%% file: main.tex
\documentclass[twocolumn, resetfootnote]{aastex7}

\usepackage{amsmath}
\usepackage[utf8]{inputenc}
\usepackage{xcolor}
\usepackage{cprotect}

\newcommand*{\kms}{\text{km}\,\text{s}\ensuremath{^{-1}}}
\newcommand*{\msun}{\ensuremath{\text{M}_{\odot}}}

\newcommand*{\LCDM}{\ensuremath{\Lambda}\text{CDM}}

\newcommand*{\EWHA}{\ensuremath{\text{EW}_{\text{H}_\alpha}}}
\newcommand*{\HA}{\ensuremath{\text{H}\alpha}}

\newcommand*{\sersic}{Sérsic}
\newcommand*{\logsm}{\ensuremath{\log_{10}(M_\star/\mathrm{M}_\odot)}}

\newcommand{\erinpaperiii}{Kado-Fong et al.~(2025, submitted)}

\newcommand*{\sagacolor}{purple}
\newcommand*{\sbgcolor}{orange}
\newcommand*{\isocolor}{green}

\newcommand*{\saganumber}{248}
\newcommand*{\sbgnumber}{508}
\newcommand*{\isonumber}{470}
\newcommand*{\sagasilvernumber}{356}

\newcommand*{\https}[1]{\href{https://#1}{\nolinkurl{#1}}}
\newcommand*{\http}[1]{\href{http://#1}{\nolinkurl{#1}}}

\definecolor{Burgundy}{RGB}{144,0,32}

\newcommand\reply[1]{\textcolor{black}{#1}} 

\shorttitle{The SAGA Survey.\ VI.}
\shortauthors{Asali et al.}

\graphicspath{{./}{figures/}}

\begin{document}

\title{The SAGA Survey.\ VI.: The Size--Mass Relation for Low-Mass Galaxies Across Environments} 

\input{authors}

\begin{abstract}

We investigate how Milky Way–like environments influence the sizes and structural properties of low-mass galaxies by comparing satellites of Milky Way analogs from the Satellites Around Galactic Analogs (SAGA) Survey with two control samples: an environmentally agnostic population from the SAGA background (SAGAbg) sample and isolated galaxies from the SDSS NASA--Sloan Atlas. All sizes and structural parameters are \reply{measured uniformly using \texttt{pysersic}} to ensure consistency across samples. 
We find the half-light sizes of SAGA satellites are systematically larger than those of isolated galaxies, with the magnitude of the offset ranging from \reply{0.05} to 0.12 dex (10–24\%) depending on the comparison sample and completeness cuts. This corresponds to physical size differences between \reply{85}-200 pc at $\logsm = 7.5$ and \reply{220}-960 pc at $\logsm = 10$. 
This offset persists among star-forming galaxies, suggesting that environment can influence the structure of low-mass galaxies even before it impacts quenching.  
The intrinsic scatter in the size--mass relation is lower for SAGA satellites than isolated galaxies\reply{, and the \sersic~index distributions of satellites and isolated galaxies are similar.}
In comparison to star-forming satellites, quenched SAGA satellites have a slightly shallower size--mass relation and rounder morphologies at low-mass\reply{, suggesting that quenching is accompanied by structural transformation and that the processes responsible differ between low- and high-mass satellites.} \reply{Our results show that environmental processes can imprint measurable structural differences on satellites in Milky Way–mass halos.} 

\end{abstract}



\section{Introduction} \label{sec:intro}

The size--mass relation is a fundamental scaling law in galaxy evolution. This well-established relation provides insight into the structural evolution of galaxies, as well as the physical processes that regulate their growth and morphology, and depends on both intrinsic properties and redshift \citep[e.g.,][]{mo_1998MNRAS.295..319M, van_der_wel_2014ApJ...788...28V, lange_2015MNRAS.447.2603L, van_dokkum_2015ApJ...813...23V, elbadry_2016ApJ...820..131E, mowla_2019ApJ...872L..13M, miller_2025ApJ...988..196M}.  

However, disagreement exists in the literature regarding how galaxy sizes depend on environment. Some studies report that galaxies are larger in denser environments \citep{cooper_2012MNRAS.419.3018C, lani_2013MNRAS.435..207L, bassett_2013ApJ...770...58B, delaye_2014MNRAS.441..203D, yoon_2017ApJ...834...73Y, siudek_2022A&A...666A.131S, yoon_2023ApJ...957...59Y}, while others find the opposite trend \citep{cebrian_2014MNRAS.444..682C, zhang_2019RAA....19....6Z, matharu_2019MNRAS.484..595M, chamba_2024A&A...689A..28C}, or no correlation at all \citep{huertas-company_2013ApJ...779...29H, shankar_2014MNRAS.439.3189S, kelkar_2015MNRAS.450.1246K, spindler_2017MNRAS.468..333S, wang_2020ApJ...889...37W, gu_2021ApJ...921...60G}. These discrepancies are complicated by differences in how galaxy size and environmental density are defined and measured. Furthermore, these studies span a large range in redshift and stellar mass, but the relationship between environment and size may itself evolve across redshift and mass. Recent work by \citet{ghosh_2024ApJ...971..142G} attempts to resolve this tension using $\sim3$ million Hyper Suprime-Cam galaxies at $0.3 < z < 0.7$ and $\logsm \gtrsim 8.9$, finding a positive correlation between large-scale density and size (i.e., galaxies in overdense regions tend to be larger). 

At lower stellar masses ($\logsm \lesssim 9$), the size--mass relation and its dependence on both intrinsic and environmental factors are less well studied, in terms of both local host environment as well as large scale environmental density. It is not clear whether trends observed at higher mass persist at the low-mass end, where galaxy growth and size evolution may be governed by different physical processes, such as stellar feedback \reply{from bursty star formation \citep[e.g.,][]{elbadry_2016ApJ...820..131E, emami_2021ApJ...922..217E} or environmental heating \citep[e.g.,][]{mayer_2001ApJ...547L.123M, donghia_2009Natur.460..605D, watkins_2023MNRAS.521.2012W}} rather than secular growth or mergers. \reply{Above $\logsm \gtrsim 8$, \citet{lange_2015MNRAS.447.2603L,lange_2016MNRAS.462.1470L} characterized the size–mass relation for different galaxy populations (by color, shape, morphology, etc.) using the GAMA survey. At lower masses, the}
ELVES survey \citep{carlsten_2021ApJ...922..267C, carlsten_2022ApJ...933...47C} reported a size–mass relation for low-mass satellites in the Local Volume, finding a tight relation for satellite galaxies in the range $5.5 < \logsm < 8.5$ with no strong dependence on star formation activity. \reply{They only minimally explored environmental effects, but found evidence for a positive correlation between size and environmental density. More robust constraints on how structural properties vary with environment require samples with well-characterized completeness}, as incompleteness likely biases samples toward smaller, higher surface brightness galaxies and may obscure true environmental trends. 

In this work, we leverage the most extensive sample of satellite galaxies between $7.5 < \logsm <10.0$ around Milky Way (MW) mass hosts from the Satellites Around Galactic Analogs Survey \citep[SAGA,][]{geha_2017ApJ...847....4G, mao_2021ApJ...907...85M, mao_2024ApJ...976..117M, geha_2024ApJ...976..118G, wang_2024ApJ...976..119W} to probe how MW-like environments shape the structure of low-mass galaxies. The SAGA satellite sample provides an ideal dataset for this purpose: a highly complete spectroscopic sample of low-mass satellites in consistent environments. 

We investigate the structural properties of SAGA Survey satellites in the context of two complementary samples: an environmentally agnostic sample of low-redshift galaxies found to be background to the SAGA hosts during SAGA Survey observations \citep[SAGAbg,][]{kadofong_2024ApJ...966..129K}, and isolated galaxies matched in redshift and stellar mass identified in \citet{geha_2012ApJ...757...85G} from the Sloan Digital Sky Survey NASA--Sloan Atlas \citep[SDSS NSA, ][]{blanton_2011AJ....142...31B}. 
We remeasure all properties for the galaxies in each sample, so that we are able to robustly determine any environment-driven differences. We present the galaxy samples used in this paper in Section~\ref{sec:data}, and describe our pipeline for remeasuring properties in Section~\ref{sec:measure} \reply{using an updated version of \texttt{pysersic} detailed in Appendix~\ref{sec:app-pysersic}}. We then present our results in Section~\ref{sec:structural-prop} and discuss their implications in Section~\ref{sec:discussion}. We summarize our conclusions in Section~\ref{sec:conc}. Distances used in this work are computed following SAGA DR3, assuming a flat \LCDM~cosmology with $H_0 = 70$\,\kms\,Mpc$^{-1}$ and $\Omega_\text{M} = 0.27$ \citep{mao_2024ApJ...976..117M}. 
 
\section{Sample Selection and Completeness Corrections} \label{sec:data}
Sizes are influenced by a variety of confounding factors, particularly at the low-mass end \citep[e.g.,][]{elbadry_2016ApJ...820..131E, wang_2020ApJ...889...37W, du_2024A&A...686A.168D, klein_2024MNRAS.532..538K, mercado_2025ApJ...983...93M}. To disentangle these effects, we compare samples across environments to more clearly isolate any imprint of the MW-like environment on galaxy structure. We first analyze the SAGA satellite population itself, then compare SAGA satellites to matched samples of galaxies in lower-density local environments (i.e. far from massive neighbors). We describe the SAGA satellite sample in Section~\ref{sec:saga-sample}, an environmentally averaged sample from SAGAbg in Section~\ref{sec:sbg-sample}, and an isolated galaxy sample from SDSS NSA in Section~\ref{sec:iso-sample}. For each sample, we describe our sample selection and the corresponding completeness correction. 

\begin{figure}[tbp]
   \centering
   \includegraphics[width=\linewidth,clip]{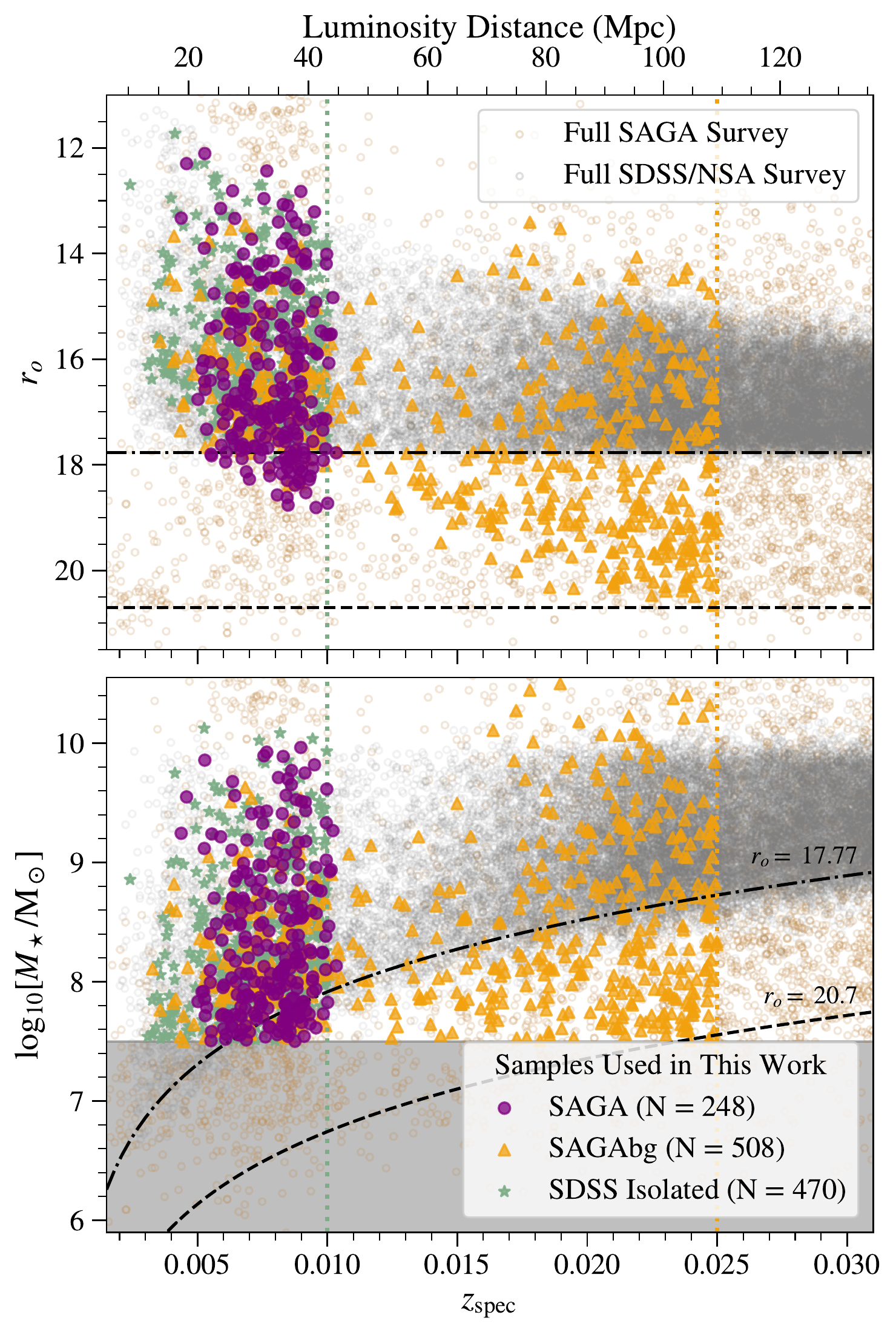}
   \caption{Distribution of $r$-band apparent magnitude ($r_{o}$, \textit{top}) and stellar mass (\logsm, \textit{bottom}) as a function of spectroscopic redshift ($z_\mathrm{spec}$) for SAGA \textit{Gold} satellites (\sagacolor, circles), SAGAbg environmentally averaged galaxies (\sbgcolor, triangles), and SDSS/NSA isolated galaxies (\isocolor, stars). Luminosity distance is shown for reference as the upper $x$-axis. The black \reply{dashed and dot-dashed} lines in both panels represent magnitude limits of $r_o = 20.7$ and $r_o = 17.77$, indicating approximate spectroscopic limits for the SAGA and SDSS surveys, respectively. The vertical dotted lines highlight the redshift criteria for the SAGAbg and SDSS/NSA isolated samples. We employ a redshift cutoff of $z<0.025$ for the SAGAbg sample (\sbgcolor~dotted line) and $z<0.01$ for SDSS/NSA isolated galaxies (\isocolor~dotted line). These redshift ranges are selected such that the samples are stellar mass-matched to our SAGA satellite sample (above $\logsm = 7.5$).}
 \label{fig:samples}
\end{figure}

\subsection{SAGA Satellite Sample} 
\label{sec:saga-sample}

The SAGA Survey\footnote{\https{sagasurvey.org}\label{fn:saga}} provides the largest systematically selected spectroscopic sample of low-mass satellite galaxies in a MW-like environment \citep{geha_2017ApJ...847....4G, mao_2021ApJ...907...85M, mao_2024ApJ...976..117M, geha_2024ApJ...976..118G, wang_2024ApJ...976..119W}. The SAGA Survey Data Release 3 \citep[DR3;][]{mao_2024ApJ...976..117M} sample includes 378 confirmed satellite galaxies in 101 MW-like systems over the distance range 25--40.75\,Mpc. Potential satellite candidates within the virial volume of each MW-analog host (300\,kpc in projected radius) were spectroscopically targeted during the survey in order to identify satellite galaxies between $6.5 \lesssim \logsm \lesssim 10$. For the purposes of this work, we primarily focus on satellites with $\logsm \ge 7.5 $ (the \textit{Gold} SAGA sample). This stellar mass limit represents the highest completeness regime of the SAGA survey for both star-forming and quenched satellites, containing a total of \saganumber~satellites in the \textit{Gold} sample that account for approximately 94\% of the expected true satellite population. Furthermore, we focus on this mass range because it aligns with the regime where our matched samples are most complete. However, since the SAGA satellite sample extends down to $\logsm \ge 6.75 $ with a well-characterized completeness correction (the \textit{Silver} SAGA sample, containing \sagasilvernumber~satellites), we also report the size--mass relation for the SAGA satellite sample alone down to this limit.

The SAGA Survey was specifically designed to maintain high spectroscopic completeness for likely satellite candidates. Nevertheless, spectroscopic incompleteness remains due to both failed candidate redshifts and candidates that were not targeted. As described in \citet{mao_2024ApJ...976..117M}, the likelihood that any galaxy is a satellite is determined through a probabilistic model that estimates the satellite rate of occurrence as a function of photometric properties. Each satellite candidate is assigned a probability of being a true satellite, $p_\mathrm{sat}$. To account for incompleteness, we measure properties for all satellite candidates that have $p_\mathrm{sat} > 0.5$, and incorporate the probability-weighted measured properties for all candidates with stellar mass $\logsm \ge 6.75$ at the assumed distance of the host into distributions and fit relations. We have measured properties for all candidates between $0.25 < p_\mathrm{sat} < 0.5$ and find that including these lower-probability candidates does not qualitatively change our results. We choose to adopt a threshold of $p_\mathrm{sat} > 0.5$ to avoid introducing sources that are unlikely to be true satellites. \reply{A detailed assessment of the effect of incompleteness on our results is presented in Section~\ref{sec:incompleteness-tests}}.

We use the same ``quenched" criteria outlined in \citet{geha_2024ApJ...976..118G}, where a satellite is considered quenched when the \HA~emission is not significant $(\EWHA - \sigma_{\EWHA})  < 2\,\mbox{\AA}$), and the specific SFR in NUV is below $-11\,\msun\,\text{yr}^{-1} $ \citep{geha_2024ApJ...976..118G, mao_2024ApJ...976..117M}. \reply{Since the size--mass relation is known to depend on galaxy morphology and star formation activity, we use star-forming and quenched classifications to analyze these populations separately in Sections~\ref{sec:size--mass} and~\ref{sec:ellip_ser_saga}. }

\subsection{SAGAbg Environmentally Averaged Sample}
\label{sec:sbg-sample}

The SAGA Survey obtained tens of thousands of redshifts in order to discover satellites around 101 MW analog hosts \citep{mao_2024ApJ...976..117M}. The vast majority of these redshifts were galaxies in the background of the SAGA host systems. The SAGAbg sample introduced in \citet{kadofong_2024ApJ...966..129K} consists of these background galaxies. This sample serves as an environmentally averaged population, as the SAGAbg galaxies were selected solely for being outside SAGA host volumes---they may be satellites of hosts more or less massive than the MW, or galaxies in isolation. However, since SAGA hosts were chosen to avoid any obvious background clusters, SAGAbg galaxies are on average in lower-density environments than SAGA satellites: 77\% of the SAGAbg galaxies in our sample lie more than 300 kpc in projected distance and $c|z - z_{\rm massive}| > 1000\ \kms{}$ from their nearest massive neighbor. The SAGAbg galaxies were observed as part of the SAGA Survey, so they were targeted and spectroscopically confirmed using the same observational strategy and instrument suite as the SAGA satellites. Therefore, the SAGAbg sample follows a similar spectroscopic completeness. 

We construct a SAGAbg matched sample to compare with the SAGA satellite sample by first identifying a redshift limit where our sample will be largely complete above $\logsm = 7.5$. The SAGA Survey's nominal apparent magnitude limit is $r_o=20.7$ \citep[indicated by the annotated black dashed lines in Figure~\ref{fig:samples};][]{mao_2024ApJ...976..117M}, although galaxies fainter than this were occasionally targeted. \citet{kadofong_2024ApJ...966..129K} estimate an effective SAGAbg limit of approximately $r_o \sim 21$. Using the stellar mass calibration described in Section~\ref{sec:measure-sm} and assuming a typical color of $(g - r)_0 = 0.45$, we find $r$-band magnitude limits of $r_o=20.7$ and $r_o=21$ correspond to a limiting stellar mass of $\logsm = 7.5$ at $z = 0.0235$ and $z = 0.027$, respectively. We therefore adopt a limit between these values of $z < 0.025$ as our redshift threshold for selecting SAGAbg galaxies. This redshift threshold maximizes the sample size while ensuring that the required completeness correction remains well-characterized. We assume that any redshift evolution in structural properties is negligible between the SAGA satellite median redshift of $z = 0.008$ and our SAGAbg limit of $z < 0.025$, which is well below the range where redshift evolution is significant \citep{kadofong_2024ApJ...966..129K}. We test the effect of shifting this redshift threshold (between $0.01 < z < 0.035$) and find that our results remain qualitatively unchanged. 


\begin{figure*}[tbp]
   \centering
   \includegraphics[width=\linewidth,clip]{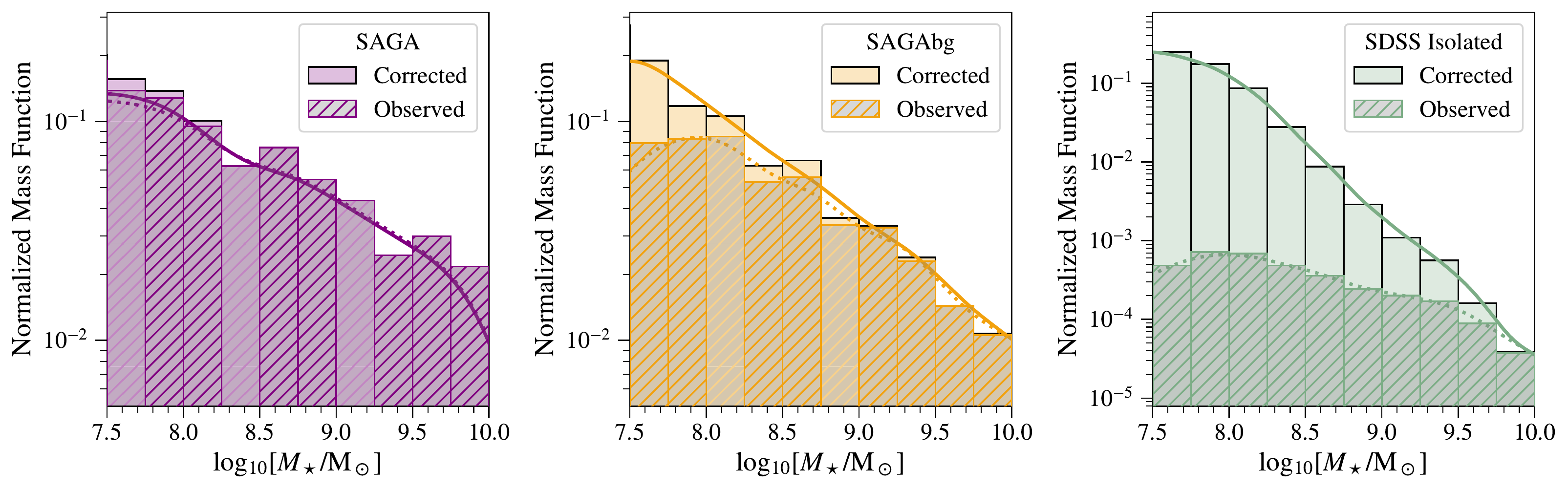}
   \caption{Stellar mass distributions for the observed and completeness-corrected samples of SAGA satellites (first panel, \sagacolor), SAGAbg galaxies (second panel, \sbgcolor), and SDSS/NSA isolated galaxies (third panel, \isocolor). The hatched histograms represent the observed (uncorrected) distributions, while the solid filled histograms show the completeness-corrected distributions. Dotted lines indicate Gaussian kernel density estimates (KDEs) for the observed distributions, and solid lines indicate KDEs for the completeness-corrected distributions. Given the large completeness correction required for the SDSS/NSA sample, we verify in Section~\ref{sec:incompleteness-tests} that incompleteness cannot fully account for our results.}
 \label{fig:mass-hists}
\end{figure*}


We include a spectroscopic completeness correction based on the effective observable volume for the SAGAbg sample using the method developed in \erinpaperiii, which we describe in brief here\footnote{We use the spectroscopic completeness correction explained in \erinpaperiii, not the photometric completeness correction}. The traditional $1/V_\mathrm{max}$ method introduced in \citet{schmidt_1968ApJ...151..393S} accounts for survey incompleteness by weighting each source by the inverse of the maximum comoving volume over which it can be observed. To account for the fact that a magnitude-limited survey can observe higher-luminosity objects out to larger distances, brighter galaxies receive smaller weights and fainter galaxies are weighted more heavily. The effective volume correction modifies this approach by replacing the maximum volume with an effective volume, computed based on the probability that a given galaxy would be successfully observed in the survey as a function of its properties, including the $r$-band apparent magnitude, $r$-band surface brightness, and $(g-r)$ color. For the SAGAbg sample, this probability is empirically derived from the fraction of targeted galaxies with successful redshift measurements for a given set of properties, and this probability is integrated over redshift to compute the effective volume. This effective volume excludes the volume corresponding to each SAGA host, as objects within $R_\mathrm{vir}$ of a host would not be included in the SAGAbg sample, but instead selected as SAGA satellites. For more details, see \erinpaperiii. For the purposes of this work, we adopt a similar effective volume correction, but our successful redshift probability evolves only as a function of $r$-band apparent magnitude and $r$-band surface brightness, since the change in these properties over our considered redshift range dominates over the change in $(g-r)$ color. 

Our SAGAbg sample consists of \sbgnumber~galaxies at $z < 0.025$. We consider our full SAGAbg sample environmentally agnostic, as galaxies are not selected by environment. However, we also define environment classifications following criteria adapted from \citet{geha_2012ApJ...757...85G} and \erinpaperiii~to construct SAGAbg subsets. Galaxies are considered \textit{isolated} if the projected 2D distance to the nearest massive galaxy exceeds 1.5 Mpc ($d_{\mathrm{massive}} > 1.5$ Mpc) and the velocity separation exceeds 1000 \kms ($c|z - z_{\rm massive}| > 1000\ \kms$). Massive galaxies are defined as those with $M_K < -23$, corresponding to a stellar mass of $\log (M_\star/M_\odot) \gtrsim 10.4$. \reply{This limit is chosen to match lower limit of $K$-band luminosity threshold used to identify MW-analogs in SAGA, originally specified in \citet{geha_2017ApJ...847....4G}.} Applying this definition, we identify \reply{69 galaxies (14\%)} in the SAGAbg sample as isolated. We also define a broader \textit{field} subset, requiring $d_{\rm massive} > 1$ Mpc and $c|z - z_{\rm massive}| > 275\ \kms$, yielding a sample of 179 galaxies (35\%). In our analysis, we primarily compare against the full SAGAbg sample out to $z < 0.025$ due to its larger size, but we also verify our results using the SAGAbg isolated and field subsets. 

\subsection{SDSS NSA Isolated Galaxies}
\label{sec:iso-sample}

We construct an isolated galaxy sample using redshifts from the NASA--Sloan Atlas \citep[NSA, ][]{blanton_2011AJ....142...31B}, which is based on SDSS DR8, imposing a redshift upper limit of $z < 0.01$ to match the redshift distribution of the SAGA satellite sample. We note that redshift-based distance estimates can carry uncertainties \reply{that may reach several hundred to over a thousand kpc} due to spectroscopic errors, scatter in the Hubble relation, and line-of-sight distortions from peculiar velocities \citep{dlr_2023ApJ...951...52D}. While such effects may influence our environmental classifications, systematic shifts are unlikely to alter our results, and spectroscopic uncertainties are included in our error propagation. 

We use the traditional $1/V_\mathrm{max}$ correction to construct a completeness-corrected sample down to our desired mass limit of $\logsm \ge 7.5$ as done in \citet{geha_2012ApJ...757...85G}, given the SDSS spectroscopic apparent magnitude limit of $r_o\sim17.77$ \citep[indicated by the annotated black dashed lines in Figure~\ref{fig:samples};][]{strauss_2002AJ....124.1810S}. The effective volume correction we apply to the SAGAbg sample, as described in Section~\ref{sec:sbg-sample}, is equivalent to the $1/ V_\mathrm{max}$ correction in the case of a magnitude-limited sample. For each galaxy, we calculate the maximum volume ($V_\mathrm{max}$) within which it could be observed given the survey's $r$-band magnitude limit of 17.77. The contribution of each galaxy is then weighted by the inverse of its $V_\mathrm{max}$ \citep{schmidt_1968ApJ...151..393S}.

\subsection{Sample Summary}

Figure~\ref{fig:samples} illustrates our galaxy samples: SAGA satellites (\sagacolor~circles), SAGAbg galaxies (\sbgcolor~triangles), and isolated galaxies from SDSS/NSA (\isocolor~stars). Our samples consist of \saganumber, \sbgnumber, and \isonumber~galaxies, respectively. These three samples are shown in the parameter space of dereddened $r$-band apparent magnitude ($r_o$, see Section~\ref{sec:measure-phot}), stellar mass (\logsm, see Section~\ref{sec:measure-sm}), and spectroscopic redshift ($z_\mathrm{spec}$). The top panel shows $r_o$ as a function of $z_\mathrm{spec}$. The black \reply{dot-dashed and dashed} lines indicate magnitude limits of $r_o = 17.77$ and $r_o = 20.7$, reflecting approximate constraints from the SDSS \citep{strauss_2002AJ....124.1810S} and SAGA \citep{mao_2024ApJ...976..117M} surveys, respectively. The bottom panel shows \logsm\ as a function of $z_\mathrm{spec}$. The black \reply{dot-dashed and dashed} curves in this panel translate the $r$-band magnitude limits into stellar mass space using the stellar mass calibration detailed in Section~\ref{sec:measure-sm}, assuming a color of $(g - r)_0 = 0.45$ which represents the median value across our samples. The vertical dotted lines in both panels indicate our chosen selection criteria for each sample in redshift space. SAGA systems span the distance range of 25–40.75\,Mpc ($0.004<z<0.011$). SAGAbg galaxies are selected below $z < 0.025$ (\sbgcolor~vertical dotted line), and SDSS/NSA isolated galaxies are selected below $z < 0.01$ (\isocolor~vertical dotted line). These redshift limits are imposed to ensure each sample is relatively complete to $\logsm = 7.5$ across the entire redshift range given each survey's apparent magnitude limit. 

Figure~\ref{fig:mass-hists} displays the original and completeness-corrected stellar mass distributions of all three samples above $\logsm = 7.5$. The hatched histograms represent the uncorrected distributions, while the filled histograms show the distributions after applying completeness corrections. We also include Gaussian kernel density estimates (KDEs) for both the observed (dotted lines) and completeness-corrected (solid lines) distributions to aid visual comparison. Our completeness correction is minimal for the SAGA satellite sample and moderate for SAGAbg. The correction for the SDSS/NSA isolated sample is more significant in the region $7.5 < \logsm < 8.0$. In particular, the SDSS/NSA isolated sample requires larger completeness corrections at low stellar masses, as we are probing beyond the nominal SDSS spectroscopic magnitude limit over a significant fraction of its volume (see bottom panel of Figure~\ref{fig:samples}). Throughout this work, we test our results to ensure that these completeness corrections do not dominate any differences seen between samples. 

\section{Homogeneously Measuring Properties} \label{sec:measure}

To eliminate the potential for measurement systematics to affect our cross-sample comparisons, we uniformly remeasure and homogeneously derive photometry, stellar masses, galaxy sizes, and structural properties, for the SAGA satellites, SAGAbg galaxies, and the SDSS/NSA isolated sample. Offsets in the size--mass relation on the basis of environment are expected to be small (e.g., \citealt{carlsten_2021ApJ...922..267C} estimated an offset of 0.04 dex for cluster satellites), and can easily be masked by systematic differences across methods of measuring size. To minimize such biases, we remeasure brightness profiles for all galaxies in our three samples using a consistent modeling procedure. We describe our brightness profile fitting method to measure structural properties in Section~\ref{sec:measure-prof} and photometry in Section~\ref{sec:measure-phot}. We then outline the calibrations used to derive stellar masses in Section~\ref{sec:measure-sm}.

\begin{figure*}[tbp]
   \centering
   \includegraphics[width=\linewidth,clip]{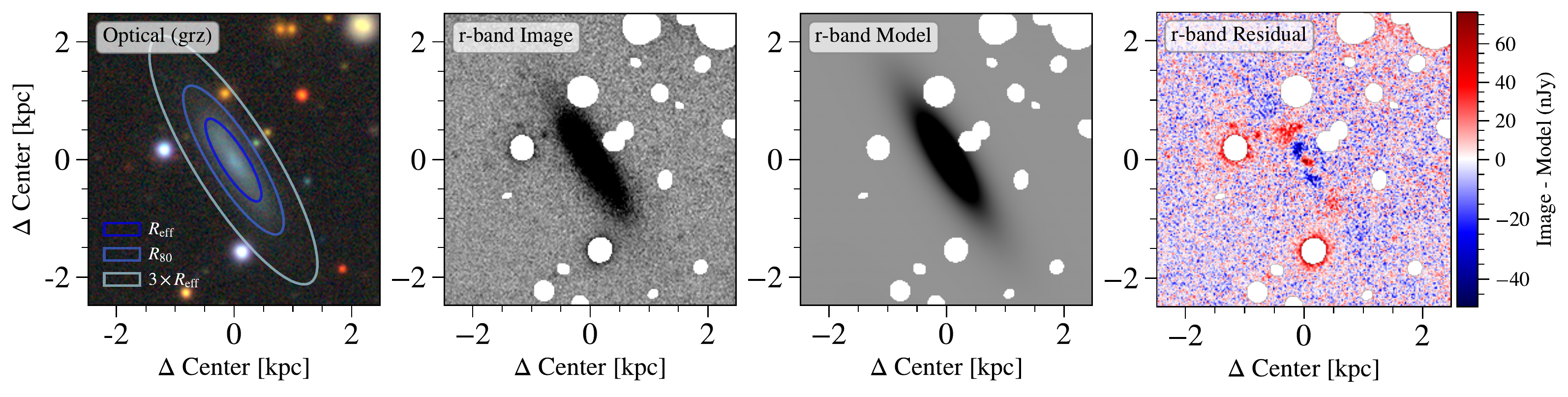}
   \cprotect\caption{Example galaxy in Legacy Survey imaging. The axes in each panel correspond to physical distances in kiloparsecs (kpc) relative to the center of the galaxy. \textit{First panel:} Optical color image ($g-$, $r-$, and $z-$bands) with \sersic~model ellipses corresponding to $R_{\mathrm{eff}}$, $R_{80}$, and $3\times R_{\mathrm{eff}}$. \textit{Second panel:} Masked $r$-band image. \textit{Third panel:} Best-fit \texttt{pysersic} model to $r$-band image. \textit{Fourth panel:} Residual between \texttt{pysersic} model image and $r$-band image.}
 \label{fig:pysersic}
\end{figure*}

\subsection{Galaxy Brightness Profiles}
\label{sec:measure-prof}

While size measurements exist for all galaxies in our samples in the DESI Legacy Imaging Surveys DR9 Tractor catalogs \citep{dey_2019AJ....157..168D, lang_2016ascl.soft04008L}, we choose not to rely on these catalog values for several reasons. First, sizes reported in the Tractor catalogs are not all obtained using a consistent model. Tractor selects from a set of five fixed models, beginning with a point source and increasing in complexity up to a single-\sersic~model, using a penalized $\chi^2$ metric that favors simpler models unless a more complex model provides a significantly better fit. This approach leads to inconsistent profile assumptions across sources. Second, the fits are performed on single-epoch imaging, which can be noisier for faint, low-mass galaxies than the co-added images. Finally, Tractor occasionally ``shreds” galaxies by misidentifying a single extended source as multiple sources, which can result in catastrophic outliers that will hamper our ability to characterize the scatter in the size--mass plane. We compare our fits with Tractor in more detail in Appendix~\ref{sec:app-tractor}. 

To ensure robust size measurements using a consistent model, we remeasure all galaxy profiles, fitting \reply{a \sersic~model to the coadded Legacy images using the Bayesian, prior-informed framework \texttt{pysersic} \citep{2023JOSS....8.5703P}. We modify the \texttt{pysersic} package to fit profiles at low \sersic~index, and outline our updates in Appendix~\ref{sec:app-pysersic}. We measure brightness profiles in $g-$, $r-$, and $z-$bands, and} choose to fit a \sersic~profile as it can reasonably approximate all galaxy types \citep{caon_1993MNRAS.265.1013C, young_1994MNRAS.268L..11Y}, allowing for reliable comparisons of sizes across our three samples. We also use these \sersic~profiles to remeasure integrated photometry and stellar masses to ensure consistency across properties as described in Sections~\ref{sec:measure-phot} and~\ref{sec:measure-sm}. 

For each galaxy, we download co-added Legacy Survey DR9 imaging from the DESI Legacy Sky Viewer\footnote{\https{legacysurvey.org}\label{fn:legacy}} in all three bands at the native pixel scale ($0.262''$/pixel), including the co-added inverse variance map and point spread function (PSF) at that location. The Legacy Survey imaging is sky pattern corrected such that the resulting images include a constant sky background, which we account for in our fitting procedure. 

We pre-process the Legacy images, masking any foreground or background sources to ensure that contaminating sources do not affect the fit of the target galaxy. To identify potential contaminant sources, we utilize the Python source extraction library \texttt{sep} \citep{2016JOSS....1...58B, 1996A&AS..117..393B}. Source detection is performed on background-subtracted images with a detection threshold set to $1.5\sigma$, where $\sigma$ represents the global background RMS. We adopt a low threshold to ensure even faint contaminating sources are identified and masked, allowing for more accurate estimation of the background during profile fitting. \reply{An example of the resulting mask is shown in the second panel of Figure~\ref{fig:pysersic}.} We crop the cutout images to 7 times the semi-major axis of the central object identified by \texttt{sep}. This choice ensures that we include enough background pixels around the galaxy for accurate fitting and avoids excessive cropping, which could compromise the fit. The pre-processed images are then manually verified to identify and correct any issues with cropping or masking. For instance, we run processing again without cropping for sources where \texttt{sep} incorrectly identifies the central object, or shift the background threshold depending on the amount of masking needed in certain fields. Approximately 15\% of galaxies across all three samples required re-processing with adjusted parameters. 

We use the \texttt{pysersic} autoprior function to generate a prior on the total flux and $R_\mathrm{eff}$ for each source. This function uses \texttt{photutils} to estimate the properties of the central object. The prior on flux is set to a wide normal distribution centered on the estimated value $f$ with a standard deviation of $2\sqrt{|f|}$. The prior on effective radius is set to a truncated normal distribution centered on the estimated value with a minimum value of 0.5 pixels. Although all cutouts are centered on the target, we set the prior on position to a normal distribution with a standard deviation of 10 pixels around the nominal position to allow for slight misalignment in the catalog coordinates. For all galaxies, we set uniform priors between $0<e<0.9$ for ellipticity, $0<\phi<2\pi$ for position angle, and \reply{$0.15<n<8.0$} for \sersic~index. \reply{In order to fit \sersic~indices below $n<0.6$, we update \texttt{pysersic} with a new approximation for the \sersic~term $b_n$ which removes the prior lower bound imposed by earlier versions. We find 58 galaxies in our sample are best fit by low-$n$ profiles. Our updated approximation is outlined in Appendix~\ref{sec:app-pysersic}}. We also fit for a flat sky background, using a prior set to a normal distribution centered on the value estimated by the \texttt{pysersic} autoprior. Finally, we run \texttt{pysersic} using the stochastic variational inference (SVI) mode to obtain posterior distributions for each parameter. We confirm that the SVI posteriors accurately reproduce the means and dispersions of the fully sampled posteriors for a subset of galaxies, consistent with the expected performance of SVI when the true parameter distributions are approximately Gaussian.

We directly compare our remeasured sizes ($r$-band half-light radii, $\log_{10}[R_{r, \mathrm{eff}}/\textrm{kpc}]$) to the Legacy Tractor catalogs, finding that 88\% of the new measurements have logarithmic residuals within 0.1 dex relative to the Tractor catalog values. There is no overall offset in our size measurements with respect to Tractor, but we find that our sizes tend to be slightly larger for compact sources and slightly smaller for extended sources compared to Tractor. This is discussed in further detail in Appendix~\ref{sec:app-tractor}. 

In addition to the Tractor catalog values, all sources in our SDSS/NSA isolated sample have size measurements from the NSA catalog \citep{blanton_2011AJ....142...31B}. We find the NSA catalog sizes have a small systematic offset relative to our measurements of $-0.02$ dex. Similarly, some fraction of our SAGA satellite (54\%), SAGAbg (36\%), and SDSS/NSA isolated (73\%) samples have existing size measurements in the Siena Galaxy Atlas (SGA-2020) \citep{moustakas_2023ApJS..269....3M}, with a systematic offset of $-0.07$ dex relative to our measurements. However, the primary purpose of this work is to identify relative differences in size across environments, and these offsets do not appear to affect our main conclusions as the systematic offset is equivalent for all three samples.  

Figure~\ref{fig:pysersic} is an example brightness profile fit to one of the SAGA satellites. The left-most panel shows the optical color image in the $g-$, $r-$, and $z-$bands with overlaid \sersic~model ellipses corresponding to $R_{\mathrm{eff}}$, $R_{80}$, and $3\times R_{\mathrm{eff}}$ from the $r$-band model shown in the third panel. We use $R_{\mathrm{eff}}$ to refer to the semi-major half-light radius so as to avoid confusion with the half-mass size, sometimes denoted as $R_{50}$. The second panel shows the $r$-band image with masked foreground and background objects and the third panel shows the best fit \texttt{pysersic} model for this $r$-band image. The right-most panel shows the residuals between the model and data images. For every galaxy, we fit all three optical bands independently. We re-measure photometry in each band as detailed in Section~\ref{sec:measure-phot} and investigate how galaxy sizes vary with wavelength across the $g-$, $r-$, and $z-$bands in Appendix~\ref{sec:lambda-r80}. 

\subsection{Photometry}
\label{sec:measure-phot}

Remeasuring the photometry is critical to ensuring consistency across all samples, as this eliminates the possibility of observational or photometric calibration mismatches influencing our results. We remeasure the total optical magnitudes using the \sersic~flux in each band and calculate dereddened magnitudes using the \citet{schlegel98} dust reddening map. We use the updated Galactic extinction coefficients for DECam filters listed on the Legacy Survey documentation\footnote{\https{https://www.legacysurvey.org/dr9/catalogs/\#galactic-extinction-coefficients}\label{fn:legacyext}}, which are calculated following the same method as \citet{Schlafly2011}. We also correct for the photometric offset between DECaLS and BASS/MzLS as outlined in \citet[Equation 2;][]{mao_2024ApJ...976..117M}. We verify that our remeasured photometry is comparable to the Legacy Tractor catalog values for the SAGA and SAGAbg sources, with a residual standard deviation of 0.23 dex.  
 
We test how much the magnitudes computed from the \sersic~profile flux differ from magnitudes computed using the aperture photometry method. Aperture photometry provides a model-independent flux estimate and serves as a check that our \sersic~fits do not systematically bias total magnitudes. 
The residuals ($r_{\mathrm{s}\acute{e}\mathrm{rsic}} - r_\mathrm{app\ phot}$) fall within an asymmetric range of $-0.1$ to $+0.3$ dex (16th--84th percentiles), with aperture magnitudes generally fainter than the \sersic-derived values. This is expected, as the \sersic~profile integrates flux out to large radii, capturing the extended low surface brightness wings of the galaxy light profile, whereas our aperture is defined at 3×$R_{\mathrm{eff}}$, which can truncate this extended emission and thus slightly underestimate the total light. Cases where aperture magnitudes are brighter typically correspond to bright, clumpy star-forming galaxies, where the \sersic~fit may smooth over substructure and miss features captured in the aperture photometry measurement. We verify that our results do not change with the use of aperture photometry measurements as opposed to \sersic~derived magnitudes. 

\subsection{Stellar Mass Calibration}
\label{sec:measure-sm}

We rederive stellar masses for all galaxies based on our remeasured photometry using the same calibration outlined in Equation 5 of \citet[][]{mao_2024ApJ...976..117M}. We assume a \citet{kroupa2001} initial mass function and an absolute solar $r$-band magnitude of 4.65 \citep{Willmer2018}. For the SAGA satellite galaxies, our stellar mass measurements are consistent with DR3 results to within 0.14 dex. There is no systematic trend in the stellar mass residuals, so we use the same stellar mass thresholds of $\logsm = 7.5$ and $\logsm = 6.75$ to define the SAGA \textit{Gold} and \textit{Silver} samples, respectively. Using these \sersic~profile based stellar masses, we include \saganumber~satellites in the Gold sample and \sagasilvernumber~in the Silver sample, compared with 243 and 360 in the DR3 \textit{Gold} and \textit{Silver} samples, respectively. 

While recent work \citep{dlr_2024arXiv240903959D} has proposed a new stellar mass calibration that is better suited to the low-mass galaxies, we opt to use the same stellar mass calibration as SAGA DR3 to ensure consistency with the completeness limit of $\logsm = 7.5$ reported in SAGA DR3. Switching to the calibration recommended by \citet{dlr_2024arXiv240903959D} results in a shift in the completeness limit in stellar mass space of approximately 0.2 dex due to differences in the underlying calibrations. However, since the relative differences between samples are the primary focus of this work, any shift in the absolute values of stellar mass is unlikely to significantly affect our conclusions.

The \citet{geha_2012ApJ...757...85G} isolated sample's original photometry and thus stellar masses come from SDSS/NSA. We remeasure these sources using Legacy imaging to ensure that any observed differences are astrophysical rather than due to inhomogeneous datasets. Our stellar masses are broadly consistent with NSA: 70\% of our measurements have stellar mass residuals within 0.2 dex relative to the NSA catalog. 

\section{Structural Properties} \label{sec:structural-prop}

With our re-measured properties in hand, we first examine the structural properties of SAGA satellites on their own, then compare them to the SAGAbg environmentally averaged and SDSS/NSA isolated galaxy samples. Section~\ref{sec:size--mass} presents the size--mass relation of SAGA satellites, and Section~\ref{sec:ellip_ser_saga} explores their ellipticity and \sersic~index distributions. Sections~\ref{sec:struct-env} and~\ref{sec:sersic-env} compare these properties with galaxies in our matched samples. We discuss our results in the context of literature size--mass relations in Section~\ref{sec:struct-lit}. 

\begin{figure}[tbp]
    \centering
    \includegraphics[width=\linewidth,clip]{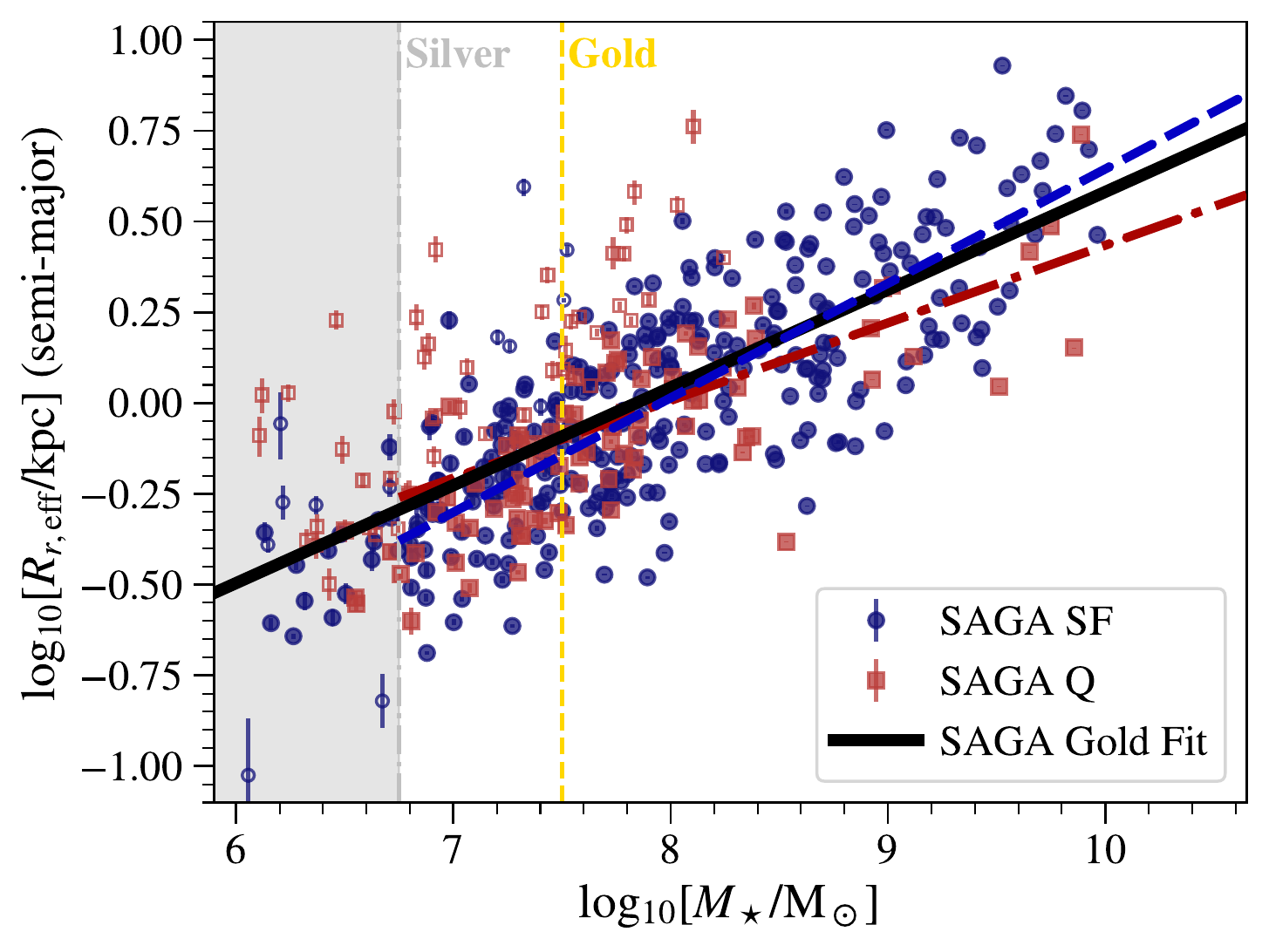}
    \caption{The size--mass relation for SAGA satellite galaxies. Star-forming (blue circles) and quenched (maroon squares) populations are both shown and fit independently. Open symbols correspond to unconfirmed candidates with high probability of being satellites ($p_\mathrm{sat} > 0.5$), with open blue circles and maroon squares indicating predicted star-forming and quenched candidates, respectively. Vertical dashed lines indicate the SAGA {\it Silver} ($\logsm = 6.75$) and SAGA {\it Gold} ($\logsm = 7.5$) mass thresholds. \reply{The black line shows our completeness corrected size--mass relation for all SAGA satellites above $\logsm = 7.5$, and the relations for star-forming and quenched satellites are shown by the blue dashed and maroon dot-dashed lines, respectively}.}
    \label{fig:size--mass-saga}
\end{figure}


\subsection{The Size--Mass Relation of SAGA Satellites} 
\label{sec:size--mass}

The size--mass relation encodes information about the physical processes driving the growth and structural evolution of galaxies. For satellites, it is likely influenced by environmental effects such as tidal interactions, ram pressure stripping, or other quenching mechanisms \citep{boselli_2014A&A...570A..69B, boselli_2014A&ARv..22...74B, wang_2020ApJ...889...37W, mercado_2025ApJ...983...93M}. 

While \citet{geha_2024ApJ...976..118G} investigated various properties of SAGA satellites, they did not measure satellite sizes. Figure~\ref{fig:size--mass-saga} shows all SAGA satellites in the size--mass plane split into quenched and star-forming satellites. We use ``size'' to refer to the semi-major $r$-band half-light radius inferred from the \sersic~profile fit, $\log_{10}[R_{r, \mathrm{eff}} / \mathrm{kpc}]$, and present our results using alternative size metrics in Appendix~\ref{sec:lambda-r80}. We find that star-forming satellites (blue circles) follow a slightly steeper relation than quenched satellites (maroon squares), and star-forming satellites tend towards larger sizes than their quenched counterparts above $\logsm > 9$. These trends are broadly consistent with the GAMA relations for early- and late-type galaxies \citep{lange_2015MNRAS.447.2603L} above $\logsm > 9$, though the GAMA relations are shallower and diverge from our measurements when extrapolated to lower masses; see Section~\ref{sec:struct-lit} for further discussion. Toward lower masses, the two populations are largely overlapping. Vertical lines indicate mass thresholds of $\logsm = 6.75 $ (SAGA \textit{Silver}) and $\logsm = 7.5$ (SAGA \textit{Gold}). 

As outlined in Section~\ref{sec:saga-sample}, we account for incompleteness in our satellite sample using the probabilistic model presented in \cite{mao_2024ApJ...976..117M}. For objects with $p_{\mathrm{sat}} > 0.5$, we re-fit their brightness profiles and assume they are at the redshift of the corresponding host galaxy to derive stellar masses and physical half-light sizes. These unconfirmed satellite candidates are shown as open points (blue circles or maroon squares for sources predicted to be star-forming and quenched, respectively) in Figure~\ref{fig:size--mass-saga}. At low masses ($\logsm < 7.5$) our incompleteness biases us toward missing larger, quenched satellites, since these objects tend to be diffuse and low surface brightness, making it more difficult to reliably measure their redshifts. As such, including this completeness correction makes the size--mass relation of quenched satellites even shallower. 

We parameterize the size--mass relation for the $r$-band effective radius in kiloparsecs in the form: 
\begin{equation}
\log_{10}(R_{r, \mathrm{eff}} / \text{kpc}) = a\logsm + b
\label{eq:size--mass}
\end{equation}
where $a$ and $b$ represent the slope and intercept, respectively. We fit the size--mass relation using weighted linear regression. A weight is assigned to each galaxy so that each mass bin contributes equally to the regression. This method of normalizing the weights in mass bins ensures that the relative completeness within each mass bin is accounted for, but the overall increase in galaxy number density towards lower stellar masses does not disproportionately impact the linear relation. Our mass-binned weighting approach ensures consistency with size–mass relations at higher masses, which are typically modeled as broken power laws with pivot masses above $\logsm > 10$, while allowing us to extend the relation to lower stellar masses. The weights of individual galaxies are given by: 
\begin{equation}
w_i = \frac{\bar{w}_i}{\sum_j \bar{w}_j}, 
\label{eq:weights-i}
\end{equation}
where $\bar{w}_i$ represents the intrinsic weights determined from our completeness correction (for SAGA, $\bar{w}_i = p_\mathrm{sat}$), and the summation over $j$ goes over all galaxies in the same mass bin as galaxy $i$. The mass bins are defined as $7.5<\logsm<10.0$, with $d\logsm=0.5$. This binning was selected since we require our mass bins to contain at least 10 data points per bin in all samples to mitigate the effect of small numbers skewing the results. We test the effects of different binning schemes, finding that our results for the size--mass relation are qualitatively consistent whether we increase or decrease the number of bins. 

We fit the size--mass relation using Equation~\ref{eq:size--mass} for all SAGA satellites, only star-forming satellites, and only quenched satellites. The black line in Figure~\ref{fig:size--mass-saga} shows our completeness corrected size--mass relation for all SAGA satellites, and the corresponding \reply{dashed and dot-dashed} lines indicate the size--mass relation for star-forming (blue) and quenched (maroon) satellites, respectively. We find the size--mass relation for all SAGA satellites above $\logsm > 7.5$ (the SAGA \textit{Gold} sample) is: 
\begin{equation}
\label{eq:saga-gold-fit}
    \log_{10}(R_{r, \mathrm{eff}} / \text{kpc}) = 0.27 \logsm - 2.11
\end{equation}
with the corresponding uncertainties being \reply{$a=0.27^{+0.023}_{-0.026}$} and \reply{$b=-2.11^{+0.225}_{-0.191}$}. We compute $\sigma$, the standard deviation of the residuals in logarithmic size at fixed stellar mass, and find \reply{$\sigma=0.21^{+0.007}_{-0.010}$.} 

We obtain uncertainties on each parameter in our reported relations ($a$, $b$, and $\sigma$) using the following method. \reply{First, to capture uncertainties due to sample selection, we generate 500 bootstrap realizations of the galaxies in each sample. Second, within each resampled set, we account for measurement uncertainties by drawing new size values from the posterior distribution of each individual galaxy's fit.} Therefore, both the uncertainties in individual galaxy fits as well as in sample selection are accounted for. \reply{For each of these 500 realization, we fit a weighted linear regression model using the weights ${w}_i$ defined in Equation~\ref{eq:weights-i}. The 16th–84th percentile range of the resulting parameter distributions is reported as the uncertainty on each parameter.}

We present the results of the full SAGA \textit{Gold} fit as well as the star-forming and quenched sample fits in Table~\ref{tab:size--mass}. We also fit the SAGA satellite size--mass relation for the \textit{Silver} sample down to $\logsm = 6.75$ where the completeness correction is still robust. We find the parameters of the size--mass relations are essentially identical to the \textit{Gold} sample, and report the results of these fits in Table~\ref{tab:size--mass} as well.

We examine how our results change when using alternative size definitions, as various size metrics may trace different physical processes in galaxies. For instance, the radius that contains 80\% of the total luminosity, $R_{80}$, is thought to be sensitive to accretion processes or halo properties at higher masses \citep[$\logsm > 9$;][]{miller_2019ApJ...872L..14M, mowla_2019ApJ...872L..13M}. We find that $R_{80}$ is consistently larger than $R_{\mathrm{eff}}$ by $\sim$0.25 dex across the full mass range. We also explore how the size--mass relation varies with wavelength. For star-forming satellites, we find the $g$- and $r$-band sizes yield nearly identical relations, while $z$-band sizes are systematically smaller. This wavelength dependence is less significant for quenched satellites. We expand on these results and show the relevant comparisons in Appendix~\ref{sec:lambda-r80}. 

To quantify the effect of incompleteness, we test how much our reported SAGA size--mass relations vary when we conservatively assume all $p_\mathrm{sat}>0.5$ objects included in our completeness correction are true satellites (e.g., set $p_\mathrm{sat}=1$ for every object). The resulting changes in the size--mass relations are minimal; the values of the parameters $a$, $b$, and $\sigma$ change by $<1.75$\% for both SAGA \textit{Gold} and \textit{Silver}. Similarly, we perform the opposite experiment and assume all $p_\mathrm{sat}>0.5$ objects are not true satellites, considering only the confirmed satellites in the SAGA sample. In this case where incompleteness is not taken into account, our results differ by $<4.5$\% and $<9.0$\% for all parameters with respect to the completeness-corrected relations above $\logsm = 6.75$ and $\logsm = 7.5$, respectively. Accounting for incompleteness corrects the size--mass relations for SAGA at the 5--10\% level.


\begin{table*}[htb]
\centering
\caption{Results of fitting the size--mass relation for different samples of galaxies. The parameters $a$, $b$, and $\sigma$ represent the slope, intercept, and intrinsic scatter. Scatter is in logarithmic size at fixed stellar mass. If you are searching for a single, environment- and morphology-independent size--mass relation for low-mass galaxies, we recommend using the bold ``All Samples Combined" relation. \label{tab:size--mass}}
\begin{tabular}{llccc}
\hline
Sample & Environment & $a$ & $b$ & $\sigma$ \\
\hline
\multicolumn{4}{c}{$7.5 \le \logsm \le 10.0 $} \\[2mm]
SAGA \textit{Gold}, All & Satellite & \reply{ $0.27^{+0.023}_{-0.026}$ } & \reply{ $-2.11^{+0.225}_{-0.191}$ } & \reply{ $0.21^{+0.007}_{-0.010}$} \\
SAGA \textit{Gold}, Star-Forming & Satellite  & \reply{ $0.30^{+0.024}_{-0.024}$ } & \reply{ $-2.41^{+0.205}_{-0.208}$ } & \reply{ $0.21^{+0.008}_{-0.009}$} \\
SAGA \textit{Gold}, Quenched & Satellite  & \reply{ $0.17^{+0.047}_{-0.052}$ } & \reply{ $-1.36^{+0.437}_{-0.386}$ } & \reply{ $0.22^{+0.021}_{-0.026}$} \\
SAGAbg & Agnostic  & \reply{ $0.24^{+0.023}_{-0.023}$ } & \reply{ $-1.87^{+0.192}_{-0.192}$ } & \reply{ $0.21^{+0.005}_{-0.007}$} \\
SDSS/NSA Isolated & Isolated  & \reply{ $0.28^{+0.030}_{-0.030}$ } & \reply{ $-2.28^{+0.246}_{-0.244}$ } & \reply{ $0.22^{+0.007}_{-0.007}$} \\
\textbf{All Samples Combined} & \textbf{Agnostic}  & \reply{ \boldmath{$0.26^{+0.016}_{-0.016}$} } & \reply{ \boldmath{$-2.10^{+0.133}_{-0.133}$} } & \reply{ \boldmath{$0.22^{+0.004}_{-0.004}$}} \\
\\
\multicolumn{4}{c}{$7.5 \le \logsm \le 9.0 $} \\[2mm]
SAGAbg Isolated & Isolated & \reply{ $0.27^{+0.058}_{-0.066}$ } & \reply{ $-2.11^{+0.542}_{-0.469}$ } & \reply{ $0.20^{+0.013}_{-0.019}$} \\
SAGAbg Field & Field  & \reply{ $0.25^{+0.067}_{-0.060}$ } & \reply{ $-2.00^{+0.478}_{-0.526}$ } & \reply{ $0.20^{+0.010}_{-0.011}$} \\
\\
\multicolumn{4}{c}{$6.75 \le \logsm \le 10.0 $} \\[2mm]
SAGA \textit{Silver}, All & Satellite & \reply{ $0.28^{+0.015}_{-0.017}$ } & \reply{ $-2.22^{+0.133}_{-0.122}$ } & \reply{ $0.21^{+0.008}_{-0.008}$} \\
SAGA \textit{Silver}, Star-Forming & Satellite & \reply{ $0.30^{+0.024}_{-0.024}$ } & \reply{ $-2.41^{+0.210}_{-0.202}$ } & \reply{ $0.20^{+0.011}_{-0.006}$} \\
SAGA \textit{Silver}, Quenched & Satellite  & \reply{ $0.19^{+0.034}_{-0.066}$ } & \reply{ $-1.49^{+0.570}_{-0.254}$ } & \reply{ $0.23^{+0.014}_{-0.032}$} \\
\hline
\end{tabular}
\end{table*}


\subsection{The Ellipticity and \sersic~Index Distributions of SAGA Satellites} 
\label{sec:ellip_ser_saga}

\begin{figure}[tbp]
    \centering
    \includegraphics[width=\linewidth,clip]{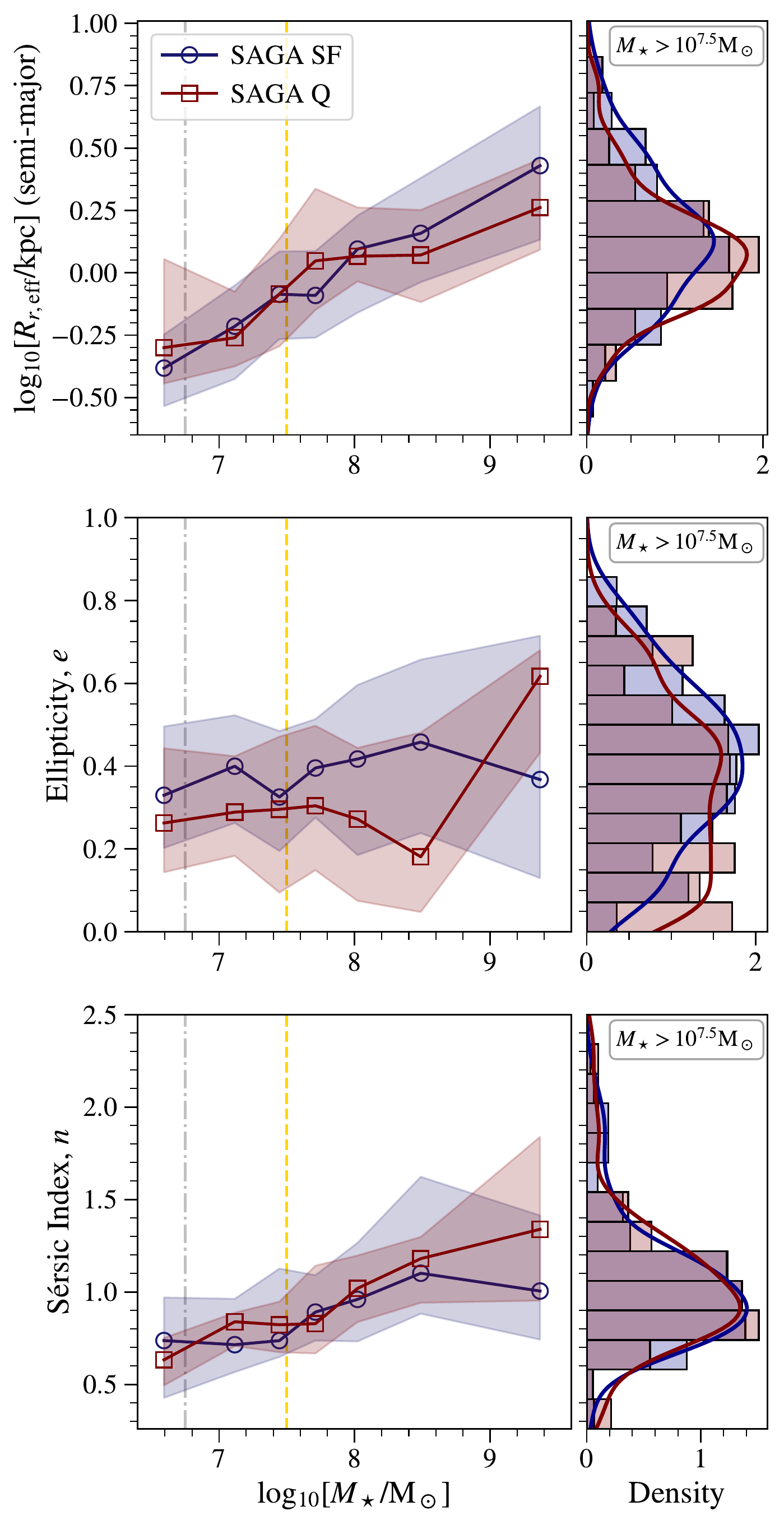}
    \caption{\reply{Half-light size ($R_{r,\mathrm{eff}}$, \textit{top}), ellipticity ($e$, \textit{middle})} and \sersic~index ($n$, \textit{bottom}) distributions of SAGA satellites. Left panels show the binned medians as a function of stellar mass, with shaded regions indicating the 16th--84th percentile ranges. Right panels show the normalized distributions for satellites in the mass range $\logsm > 7.5$. Gaussian KDE density functions are included to illustrate the overall shape of each distribution. In \reply{all} panels, distributions show the completeness corrected satellite population, split by star-forming (SAGA SF, blue) and quenched satellites (SAGA Q, maroon).
}
    \label{fig:shapes-saga}
\end{figure}
 
We next examine differences in other structural properties for SAGA satellites, namely the ellipticity and \sersic~index, which characterize a galaxy’s light distribution and elongation, respectively. Figure~\ref{fig:shapes-saga} shows both the distributions (right panels) and the median trends as a function of stellar mass (left panels) for star-forming (SF, blue) and quenched (Q, maroon) SAGA satellites. As before, to account for incompleteness, we weight the distributions by satellite probability. In the left-hand panels, we show the binned medians and 16th–84th percentile ranges for each parameter as a function of stellar mass. Bins are computed using stellar mass percentiles to ensure approximately equal numbers of galaxies in each bin. In the right-hand panels, we show the weighted histograms for only the SAGA \textit{Gold} sample, and overlay Gaussian kernel density estimates (KDEs) to visually highlight the overall trends in each population.

We find that the ellipticities of star-forming satellites are roughly Gaussian distributed around $e\sim0.4$. In contrast, the ellipticity distribution of quenched satellites tends toward lower-$e$ values. This suggests that quenched satellites are, on average, intrinsically rounder than star-forming satellites in three dimensions, which is consistent with previous findings \citep{padilla2008, kadofong_2020ApJ...900..163K, carlsten_10_2020ApJ...891..144C}. We also note the ellipticity distribution of quenched satellites exhibits a bimodal structure. We compute Hartigan’s dip test statistic \citep{hartigan} to assess the significance of this observed bimodality. This non-parametric test measures the maximum difference between the empirical distribution function and the best-fitting unimodal distribution; smaller $p$-values indicate stronger evidence against unimodality. The test yields a $p$-value of 0.05, indicating that the distribution is not well described by a unimodal model. 

We find that the lower-$e$ peak in the quenched population corresponds to lower-mass satellites, as evidenced by the drop in median ellipticity at $\logsm = 8.5$ shown in the \reply{middle} left panel of Figure~\ref{fig:shapes-saga}. That is, the trend of quenched satellites being rounder applies only to lower-mass satellites. We find quenched galaxies at moderate ellipticities ($0.3<e<0.6$) across the full mass range of $6.75 < \logsm < 10$. In contrast, the ellipticity distribution of star-forming satellites is shifted toward higher values, which again implies that they are, on average, more elongated in three dimensions. 

In contrast to ellipticity, the \sersic~index distributions appear broadly similar for both quenched and star-forming satellites. Indeed, the two are statistically consistent with being drawn from the same distribution. Most satellites, regardless of their star formation activity, have \sersic~indices close to 1, characteristic of disk-like light profiles. For both the star-forming and quenched satellite populations, the \sersic~index distribution has a small tail toward higher $n$ values, reflecting more centrally concentrated light distributions. While the \sersic~index of the star-forming population does not show a significant trend with mass, we find a moderate positive correlation ($r=0.53$) between \sersic~index and stellar mass for quenched satellites. This is reflected in the bottom left panel of Figure~\ref{fig:shapes-saga}, where the binned median \sersic~index for quenched satellites rises with increasing stellar mass. The quenched satellites that have elevated \sersic~indices in the range of 2.0--2.5 correspond to the most massive quenched satellites. This suggests that massive quenched satellites may have undergone some structural transformations that differ from the rest of the quenched satellite population. We discuss the implications of these results in further detail in Section~\ref{sec:discussion}. 

We again test the robustness of our conclusions against incompleteness by comparing results using both the uncorrected sample (including only confirmed satellites) and a maximal completeness correction (assuming all unconfirmed satellites are real). In both cases, we find the same qualitative differences between the quenched and star-forming populations. Moreover, although Figure~\ref{fig:shapes-saga} shows only the distributions for the \textit{Gold} sample, we verify that our findings remain consistent down to the SAGA \textit{Silver} mass limit. 

\subsection{The Size--Mass Relation Across Environments}
\label{sec:struct-env}

We next examine the structural properties of SAGA satellites in comparison with the SAGAbg environmentally averaged and SDSS/NSA isolated samples to probe the effect of environment on galaxy sizes. As before, we fit a weighted linear regression model in order to account for our completeness correction, using mass-normalized weights as described in Equation~\ref{eq:weights-i}. We report these completeness corrected size--mass relation fits in terms of $r$-band effective radius for all our samples in Table \ref{tab:size--mass}. For the SAGAbg sample, the intrinsic weight of the $i$\textsuperscript{th} galaxy is given by 
$\bar{w}_i = V_{\mathrm{comoving}(z_\mathrm{max})}/V_{\mathrm{eff},i}$
and for the SDSS/NSA sample the intrinsic weights are given by $\bar{w}_i = 1/V_{\mathrm{max},i}$, as described in Section~\ref{sec:data}. 

\begin{figure*}[!tbp]
    \centering
    \includegraphics[width=\linewidth,clip]{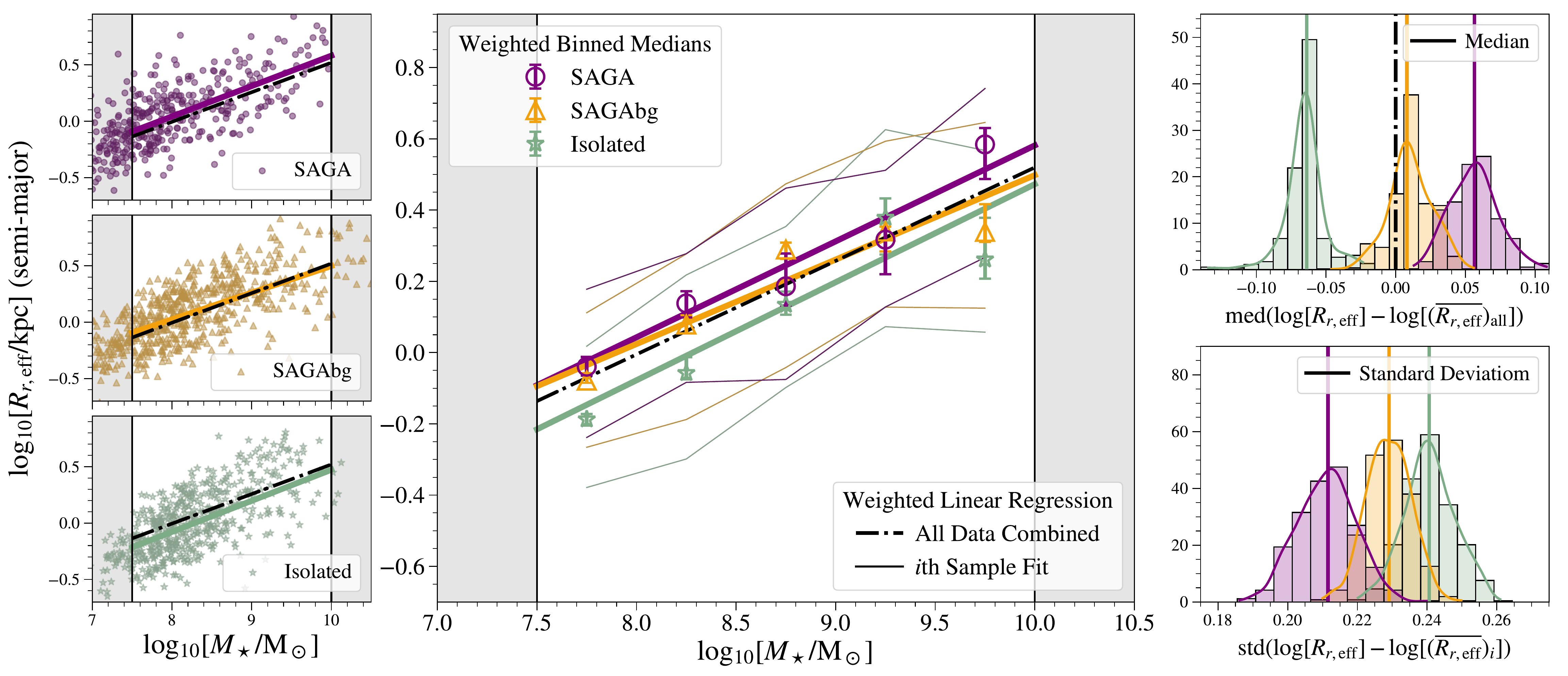}
    \caption{The $r$-band effective radius size--mass distribution for SAGA satellite galaxies (\sagacolor, \textit{top left}), SAGAbg environmentally averaged galaxies (\sbgcolor, \textit{middle left}), and SDSS/NSA isolated galaxies (\isocolor, \textit{bottom left}). \textit{Middle panel:} Weighted binned medians and size--mass relation fits for all three samples. The binned 16th-84th percentile scatter for each sample is indicated by the thin colored lines. The mass ranges outside of $7.5 < \logsm < 10.0$ are shaded grey in the left column and middle panel, as these represent regions of parameter space where the samples are significantly incomplete. Only data within that mass range is considered. Bootstrapped samples of summary statistics of the normalized-to-average sizes are plotted in the right column. The median distributions are shown for normalized-to-average size fit to all data (\textit{top right}) and the standard deviation distributions are shown for normalized-to-average size fit to each sample separately (\textit{bottom right}).}
    \label{fig:size--mass-samples}
\end{figure*}

Figure~\ref{fig:size--mass-samples} presents the size--mass relation comparing our three samples: SAGA satellites, the SAGAbg sample, and SDSS/NSA isolated galaxies. In the left column, we show the data and completeness-corrected size--mass relations for each sample independently. In the middle panel, we show the same completeness-corrected size--mass relations for all samples, along with a size--mass relation fit to all samples combined.
We show weighted binned medians for each sample using the same mass bins used in Equation~\ref{eq:weights-i} to compute our mass-normalized weights. The error bars indicate bootstrapped uncertainties on these medians. The thin lines indicate the 16th-84th percentile scatter for each sample across bins. The parameter values for all size--mass relations are included in Table \ref{tab:size--mass}. As each relation is fit to the mass-normalized data, these size--mass relations are robust across the mass range of $7.5 \le \logsm \le 10.0 $.

Across this mass range, the sizes of SAGA satellites are systematically larger than those of galaxies in both the SAGAbg sample, which is environmentally agnostic but on average probes lower-density environments than those of the SAGA hosts, and the isolated sample from SDSS/NSA. The normalization of both the SAGAbg and isolated sample size--mass relations are slightly offset toward smaller sizes, as shown in the middle panel of Figure~\ref{fig:size--mass-samples}. This offset is a statistically significant difference. To better demonstrate and quantify this difference, we can compute the normalized-to-average size distributions for each sample. 

We define the normalized-to-average effective radius as: 
$$ \log_{10}[R_{r, \mathrm{eff}} / \mathrm{kpc}] - \log_{10}[\big(\overline{ R_{r, \mathrm{eff}} }  (M_\star) \big)_{i}/ \mathrm{kpc}] $$
where $\log_{10}[\big(\overline{ R_{r, \mathrm{eff}} }  (M_\star) \big)_{i}/ \mathrm{kpc}] $ is the size derived at a given mass from the size--mass relation fit to the $i$th sample. Thus, the normalized size can be considered stellar-mass independent and probes the difference in the size--mass relation across these samples. For each sample, we compute two versions of the normalized-to-average size. First, we calculate the normalized-to-average size using one single fit to all the data (the black dot-dashed line in the middle panel of Figure~\ref{fig:size--mass-samples}). 
We denote this using the subscript `all'. We also compute the normalized-to-average size using a size--mass relation fit to each sample independently, which we denote using the subscript `$i$' and discuss these results in Section~\ref{sec:scatter}. 

Normalizing all samples to a single size--mass relation fit allows us to more easily probe the difference in sizes between the samples. To quantify this difference, we compute the weighted medians of normalized size distribution for each sample and use bootstrap sampling to quote 16th-84th percentile uncertainties on the medians. These median samples are shown for all three samples in the top right panel of Figure~\ref{fig:size--mass-samples}. In this format, the offset in the median of the size--mass distribution between the isolated, SAGAbg, and SAGA satellite samples is clearly evident. We find the median normalized size of the SAGAbg sample is \reply{$0.05$} dex lower than the median normalized size of SAGA satellites, and this offset is significantly larger than their respective median uncertainty intervals. \reply{We assess the statistical significance of this result in two ways: a two-sample Kolmogorov–Smirnov test on the median distributions yields a $D$ statistic of 0.88, indicating a highly significant difference \citep{smirnov_1948, ks_test_1951}; and a two-sample Anderson–Darling test on the same distributions similarly gives $p<0.001$, corresponding to $>3\sigma$ significance \citep{anderson_darling_1952}.} 

\reply{We further apply a Chow test to compare the underlying linear regressions directly \citep{chow_1960}. As expected, the significance is reduced relative to the median distribution comparisons due to the larger intrinsic scatter in the size--mass plane, but we still find tentative evidence that the SAGA and SAGAbg relations differ at the $\sim1.6\sigma$ level, while the SAGA and isolated galaxy relations differ at $>3\sigma$. Applying the Anderson–Darling test directly to the normalized size distributions, we recover consistent results: the normalized size distributions of SAGA and SAGAbg are marginally separable ($\sim1\sigma$), whereas SAGA and the isolated sample are clearly distinct ($>3\sigma$). Additionally, we compute Kendall’s $\tau$ correlation coefficient in the normalized size–stellar mass plane to assess whether the linear model is well specified, and find $p=0.17$, indicating no significant monotonic trend and confirming that the normalized size residuals do not depend on stellar mass \citep{kendall_1938}.} 

\reply{The median offset of 0.05 dex between the SAGA and SAGAbg samples} corresponds to a physical difference of \reply{$+85$} pc at $\logsm =7.5$ and \reply{$+220$} pc at $\logsm =10.0$. The offset between the SDSS/NSA isolated sample and SAGA satellites is even larger, at $0.12$ dex ($+200$ pc at $\logsm = 7.5$, $+960$ pc at $\logsm = 10.0$). The SDSS/NSA sample is less complete at low surface brightness, which contributes to the larger offset; however, we show in Section~\ref{sec:incompleteness-tests} that this incompleteness alone cannot account for the full effect. In other words, we find that despite the large scatter in the size--mass relation for individual galaxies, at a population level, satellites are between \reply{$\sim$10-24}\% larger in effective radius than their counterparts in lower-density or isolated environments. 

As mentioned in Section~\ref{sec:sbg-sample}, a small subset \reply{(14\%)} of our SAGAbg sample also meet our isolation criteria of $d_{\mathrm{massive}} > 1.5$ Mpc. We fit the size--mass relation to this SAGAbg isolated subset, and find this relation is also offset from the SAGA satellite relation. Given the relatively small number of sources in this SAGAbg isolated sample \reply{(69)} and our requirement of at least 10 objects per mass bin, we only fit the size--mass relation for this sample between $7.5\le \logsm \le 9.0$. Again, we find that SAGA satellites are offset towards larger sizes at a given stellar mass with respect to the SAGAbg isolated subset by $+0.05$ dex (10\%), corresponding to a physical size difference of $+85$ pc at $\logsm =7.5$ and $+220$ pc at $\logsm =9.0$. We also compare with the SAGAbg field subset, and find it is again offset toward smaller size from the SAGA satellite population at the $\sim$10\% level. While the median size of the SAGAbg isolated subset is essentially equivalent to that of the full SAGAbg sample, we note that the slope of the SAGAbg isolated size--mass relation is slightly steeper than that of the full sample ($0.27$ compared with $0.24$), and matches the slope of the SAGA satellite sample (\reply{$0.27$}). 

\begin{figure}[!tbp]
    \centering
    \includegraphics[width=\linewidth,clip]{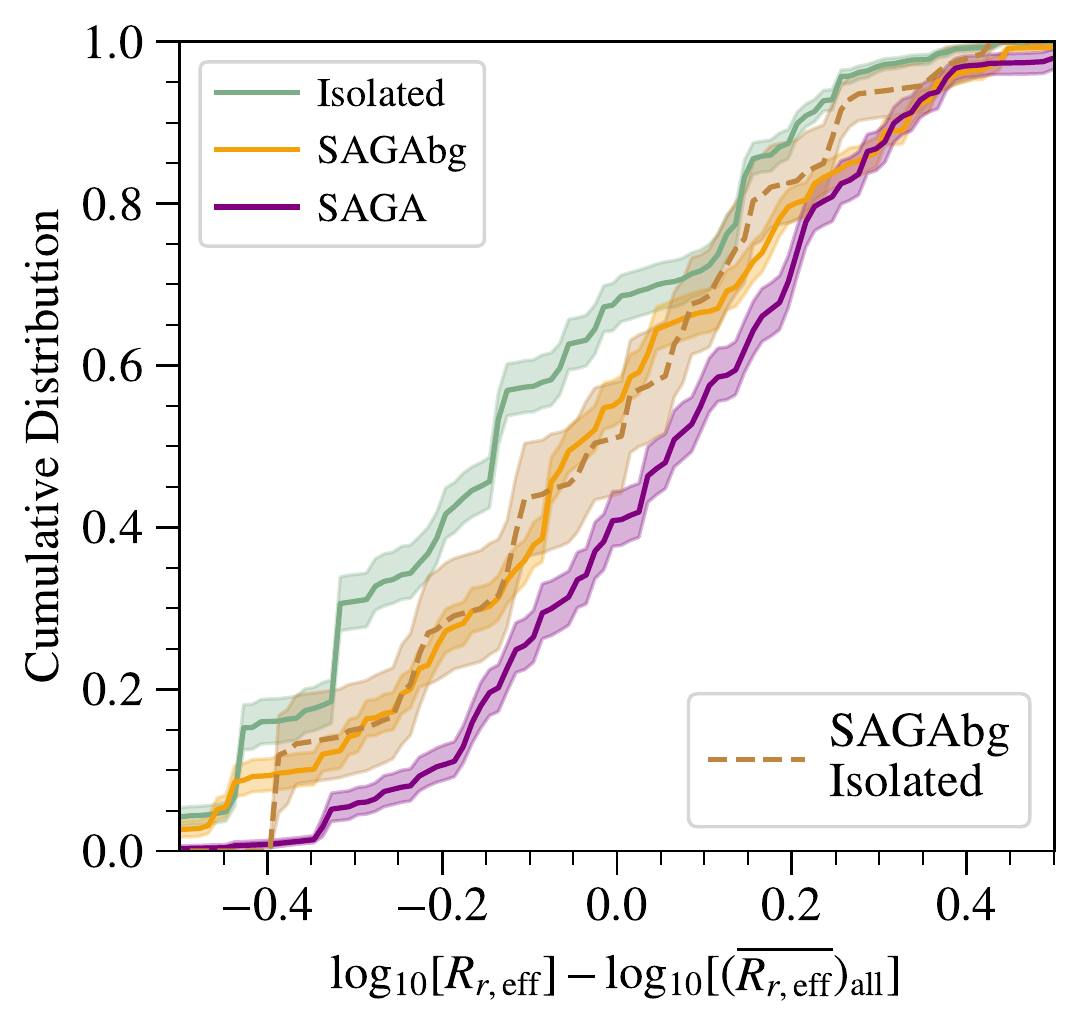}
    \caption{Cumulative distributions of the normalized-to-average size for the SAGA satellite galaxies (\sagacolor), SAGAbg environmentally agnostic galaxies (\sbgcolor), and SDSS/NSA isolated galaxies (\isocolor) in the mass range $7.5 < \logsm < 10.0$. Data is normalized to the ``all data combined" fit shown by the dot-dashed line in Figure~\ref{fig:size--mass-samples} and in Table~\ref{tab:size--mass}. We also include the CDF for the SAGAbg isolated (dashed, brown) subset. 
    }
    \label{fig:sbg-iso}
\end{figure}

This difference in slope reflects a change in the shape of the normalized-to-average size distributions between the full SAGAbg sample and the SAGAbg isolated subset. Figure~\ref{fig:sbg-iso} shows the cumulative distributions of normalized-to-average size for the SAGAbg isolated subset ($d_{\mathrm{massive}} > 1.5$ Mpc), as well as the main SAGAbg sample, the SAGA satellite sample, and the SDSS/NSA isolated sample. All distributions have been normalized to the ``all data combined'' fit provided in Table~\ref{tab:size--mass}. The SAGAbg isolated subset follows the full SAGAbg sample, and is offset from the SAGA satellite population at the $\sim$10\% level. 
However, while the full SAGAbg sample overlaps with the SAGA satellite sample at the high-size end (i.e., some SAGAbg galaxies are comparably large for their mass as the largest satellites), applying the isolation cut results in a change in the shape of the cumulative distribution. These larger-than-average galaxies are present in the full SAGAbg sample, and are removed by the isolation criteria, reinforcing that the SAGA satellites are larger on average than their counterparts in lower density environments. 

\begin{figure}[!tbp]
    \centering
    \includegraphics[width=\linewidth,clip]{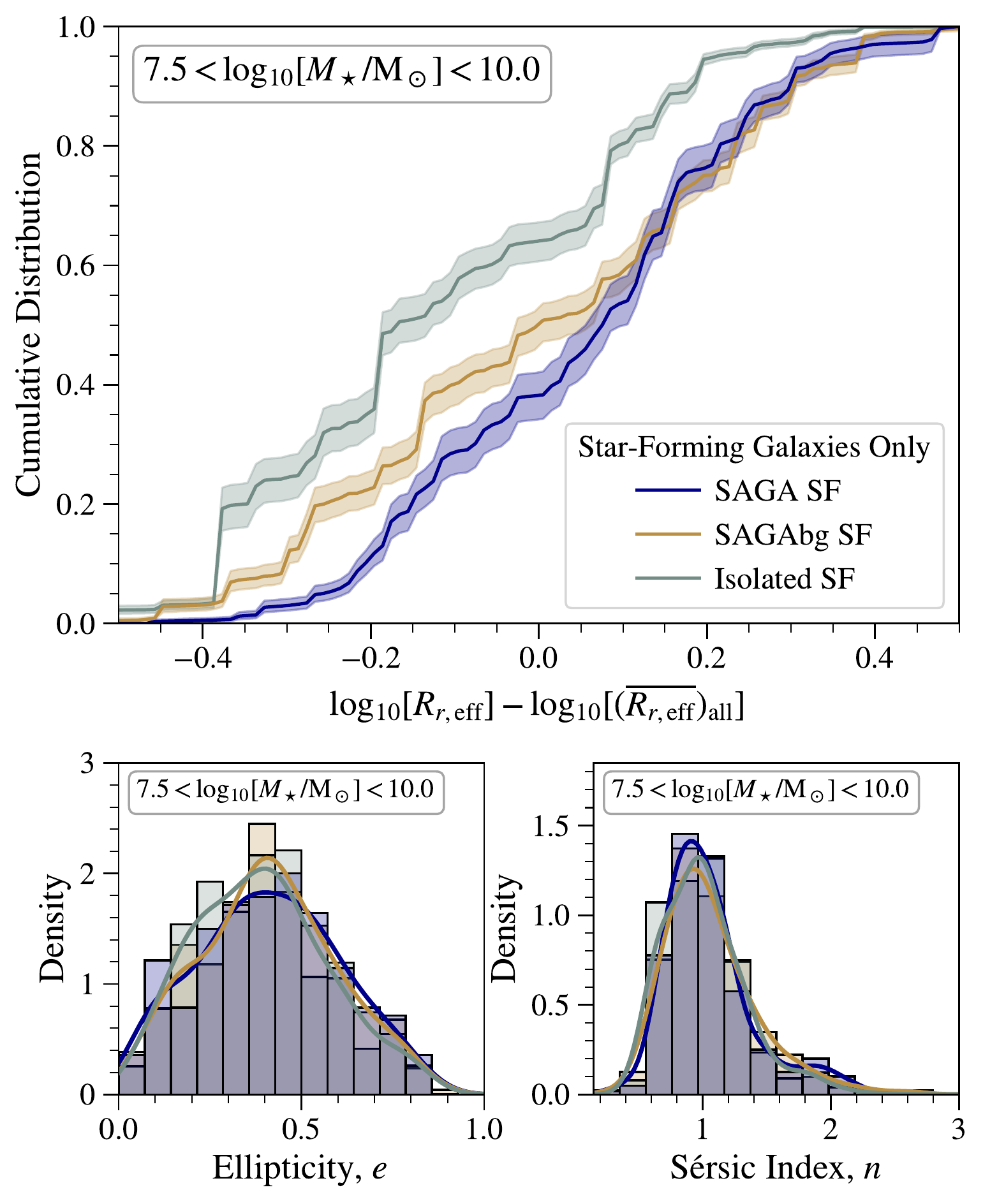}
    \caption{
    \textit{Top:} Cumulative normalized size distributions for star-forming galaxies only, across the SAGA satellites (dark blue), SAGAbg galaxies (muted \sbgcolor), and isolated galaxies (muted \isocolor). 
    \textit{Bottom Left:} Ellipticity ($e$) distributions for star-forming galaxies in each sample. 
    \textit{Bottom Right:} \sersic~index ($n$) distributions for the same samples. 
    }
    \label{fig:smr-sf-only}
\end{figure}

While the fitted relations imply a uniform offset across stellar mass, the binned medians reveal that the offset is most pronounced at both the low- and high-mass ends of the distribution. At the low-mass end ($7.5 < \logsm < 8.5$), one might assume this offset is driven solely by the presence of quenched objects in the SAGA satellite sample, which we have shown in Section~\ref{sec:size--mass} tend to be larger at low mass. However, restricting to star-forming galaxies only, low-mass SAGA satellites remain systematically larger than their counterparts in lower-density environments. This is illustrated in Figure~\ref{fig:smr-sf-only}, which shows the cumulative distribution of the normalized-to-average size for only star-forming galaxies in all three samples. The median normalized-to-average size of star-forming SAGA satellites is $0.03$ dex above that of the SAGAbg star-forming subset, but the two distributions have different shapes and the SAGA satellite sample has fewer galaxies with smaller average sizes. The median offset is smaller than the \reply{$0.05$} dex offset between the full samples shown in Figure~\ref{fig:size--mass-samples}, but this difference can be understood by the fact the SAGA satellite sample has a larger fraction of low-mass quenched galaxies than the full SAGAbg sample, accounting for the additional offset and shifting the distribution shape. Additionally, star-forming SAGA satellites remain on average $0.1$ dex larger than star-forming isolated counterparts from SDSS/NSA, as the full isolated sample is predominately composed of star-forming galaxies. We cannot draw strong conclusions about environmental effects on high-mass quenched galaxies, since our samples of these galaxies are relatively small: 7 SAGA satellites, 35 SAGAbg galaxies, and only 3 isolated galaxies are high-mass ($\logsm > 9.0$) and quenched. However, we note that the median size of these massive quenched isolated galaxies is smaller than that of quenched SAGA satellites. We discuss the implications of this mass-dependent offset and its connection to star-forming properties in Section~\ref{sec:discussion}. 

Finally, we examine how alternative size definitions and wavelength-dependent trends affect the size--mass relation across environments. Further details and comparisons are presented in Appendix~\ref{sec:lambda-r80}. As with the SAGA satellite sample, we find that $g$- and $r$-band sizes are consistent, while $z$-band sizes are systematically smaller for both the SAGAbg and isolated samples. We also find that $R_{80}$ remains uniformly offset from $R_{\mathrm{eff}}$ by $\sim$0.25 dex across all three samples. 

\subsection{Scatter in the Size--Mass Relation}
\label{sec:scatter}

In addition to measuring the offset in the size--mass relation as a function of environment, we also investigate the intrinsic scatter in the size--mass relation. The intrinsic scatter may reflect a range of physical processes, including scatter in the stellar–halo mass relation \citep{wang_2020ApJ...889...37W}, variations in the specific angular momentum of baryonic matter \citep{du_2024A&A...686A.168D}, or stellar migration driven by bursty star formation \citep{elbadry_2016ApJ...820..131E}.
To quantify the scatter in the size--mass relation, we now calculate the normalized-to-average sizes for each sample using fits to each sample independently, rather than normalizing to one single fit. This method allows us to probe the intrinsic scatter in the size--mass relation for each environment separately. As before, we bootstrap sample from these normalized size distributions for each sample and compute weighted standard deviations to measure the difference in the intrinsic scatter for each sample. These bootstrap samples are shown in the bottom right panel of Figure~\ref{fig:size--mass-samples}. We find that the size--mass relation for SAGA satellites has a marginally smaller intrinsic scatter than that of isolated galaxies, \reply{implying that if bursty star formation is the primary driver of scatter in the size--mass plane, satellites experience either less frequent or lower-amplitude bursts compared with isolated galaxies.}

\subsection{Quantifying Incompleteness Effects and Validating Results}
\label{sec:incompleteness-tests}

\begin{figure}[!tbp]
    \centering
    \includegraphics[width=\linewidth,clip]{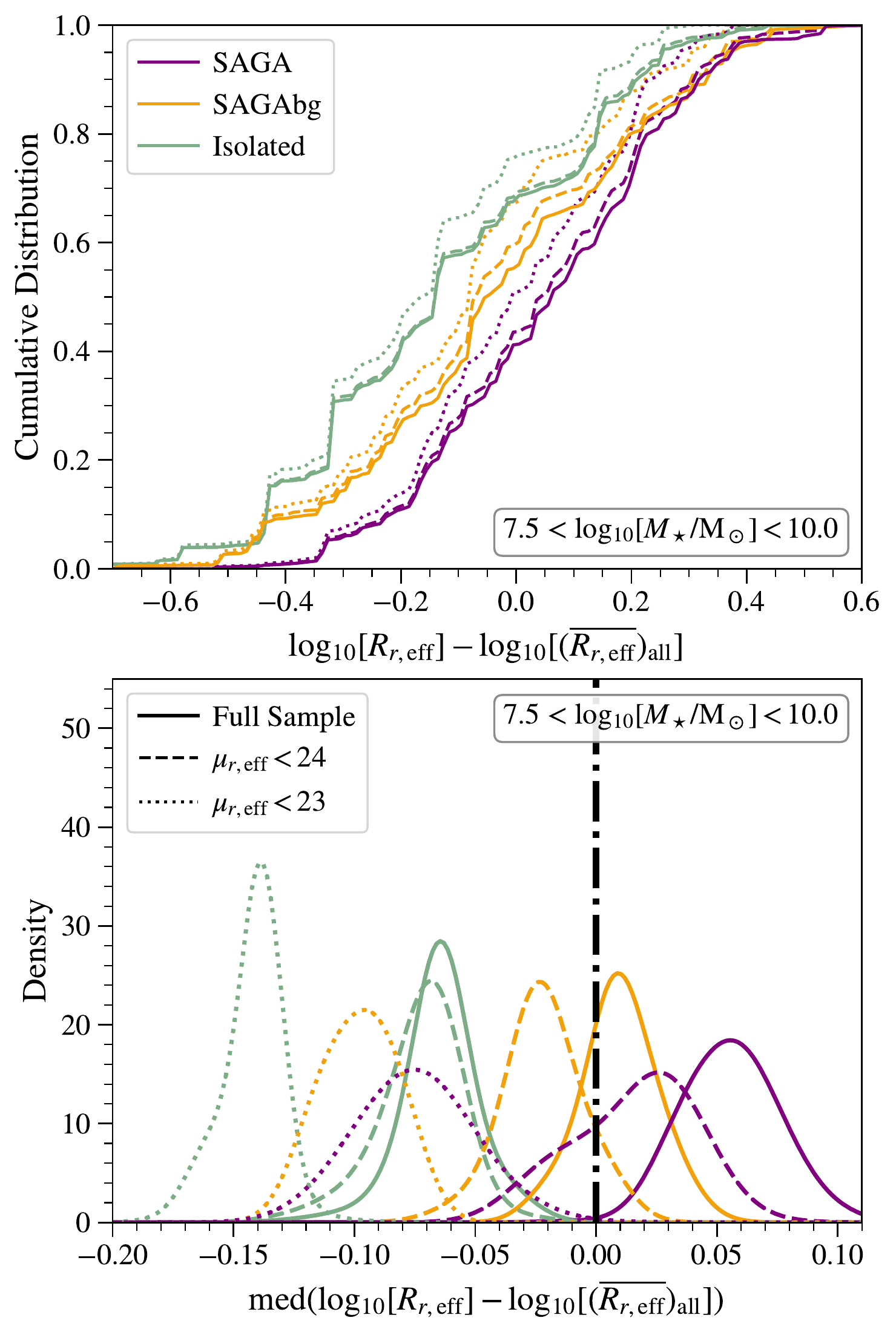}
    \caption{\reply{We test the effects of incompleteness on our results by measuring the observed offsets using surface brightness thresholds.} \textit{Top:} Cumulative distributions of the normalized-to-average size for the SAGA satellite galaxies (\sagacolor), SAGAbg environmentally agnostic galaxies (\sbgcolor), and SDSS/NSA isolated galaxies (\isocolor) in the mass range $7.5 < \logsm < 10.0$, as shown in Figure~\ref{fig:sbg-iso}. Two additional CDFs are included per sample where surface brightness cuts of $\mu_{r, \mathrm{eff}} < 23 \ \mathrm{mag \ arcsec}^{-2}$ (dotted lines) and $\mu_{r, \mathrm{eff}} < 24 \ \mathrm{mag \ arcsec}^{-2}$ (dashed lines) are applied. \textit{Bottom:} Bootstrapped samples of the median normalized-to-average size for each subset shown in the top panel. The solid lines in the right panel are equivalent to the distributions shown in the top right panel of Figure~\ref{fig:size--mass-samples}. Even when restricting the samples to the lower surface brightness sources where completeness is highest, the median size difference between the isolated and satellite samples persists.}
    \label{fig:norm-size-cdf}
\end{figure}

A possible concern is whether incompleteness in the SDSS/NSA sample toward low surface brightness galaxies could be driving the observed offset between these two samples. SDSS aims to be complete to an average surface brightness of $\mu_{r, \mathrm{eff}} \sim 24.5 \mathrm{\ mag \ arcsec}^{-2}$ \citep{strauss_2002AJ....124.1810S, shen_2003MNRAS.343..978S}, and we apply a $1/V_\mathrm{max}$ correction to the isolated sample under the assumption of magnitude-limited incompleteness. However, it is possible surface brightness selection effects would lead to the omission of large, diffuse galaxies in the isolated sample. This could artificially decrease the median size of the isolated sample as compared with SAGA, where low surface brightness objects are included. 

As an initial test, we compare the normalized-to-average size distributions of all three samples after applying surface brightness cuts of $\mu_{r, \mathrm{eff}} < 24$ and $< 23 \ \mathrm{mag \ arcsec}^{-2}$. Figure~\ref{fig:norm-size-cdf} shows how the cumulative distributions of normalized-to-average sizes for the SAGA satellite, SAGAbg, and isolated samples change when applying these two surface brightness cuts (top panel). As expected, the cumulative distributions shift leftward in both cases, indicating the removal of larger-than-average galaxies at a given mass. However, the offset between samples remains: the median size difference between SAGA satellites and isolated galaxies is 0.095 dex (19.6\%) and 0.059 dex (12.7\%), for surface brightness limits of $24 \mathrm{\ mag \ arcsec}^{-2}$ and $23 \mathrm{\ mag \ arcsec}^{-2}$, respectively. This is also evident in the lower panel of Figure~\ref{fig:norm-size-cdf}, which shows the median distributions for each subset; all three distributions shift towards lower median values of normalized size and the median offset between the samples persists, although it decreases from a value of 0.12 dex to 0.059 dex. These results imply that while surface brightness incompleteness contributes modestly to the observed offset, it cannot fully account for it. Even within the most complete, high surface brightness subsets, SAGA satellites remain systematically larger.

To further assess the impact of surface brightness incompleteness, we perform a test designed to quantify the bias this incompleteness (and other observational limitations) could induce in the measured size offset between samples. Specifically, we use SAGA fields that overlap with the SDSS/NSA footprint to evaluate how many galaxies are included in the full SAGA spectroscopic sample but missed by NSA above its limiting magnitude of $r < 17.77$. We construct 500 mock NSA-like samples by randomly selecting 600 galaxies from the SAGA catalogs and removing any that are missing from the NSA catalog, in order to approximately match the size of the SDSS/NSA isolated sample considered in this work. For each mock sample, we compute the normalized size offset relative to the full SAGA ``true" sample using the same weighting scheme and methodology as described before. As expected, this incompleteness biases the SDSS/NSA sample to smaller sizes, but we find that in fewer than 0.1\% of realizations does the induced offset exceed 0.08 dex, and never reproduces the observed 0.12 dex offset between SAGA satellites and the SDSS/NSA isolated sample. On average, this incompleteness bias leads to a median offset of only $\sim$0.025 dex, suggesting that surface brightness incompleteness alone cannot explain the observed size difference.

\reply{We similarly test the robustness of our results to incompleteness in the SAGA satellite sample by repeating our analysis using different thresholds on satellite membership probability, $p_{\mathrm{sat}}$. In addition to our fiducial completeness-weighted analysis using $p_{\mathrm{sat}} > 0.5$, we explore thresholds ranging from $p_{\mathrm{sat}} > 0.25$ to $p_{\mathrm{sat}} > 0.7$. We find that the observed trends persist across all thresholds, and our results are unchanged within uncertainties. When using the most inclusive set ($p_{\mathrm{sat}} > 0.25$, comprising 265 satellite candidates), we recover a median size offset of 0.048 dex relative to the SAGAbg sample and 0.120 dex relative to the isolated sample. Restricting to the highest-confidence set ($p_{\mathrm{sat}} > 0.7$, comprising only 28 satellite candidates), we recover a median size offset of 0.046 dex relative to the SAGAbg sample and 0.119 dex relative to the isolated sample. Our results remain unchanged even when no completeness weighting is applied (equivalent to setting $p_{\mathrm{sat}} = 1.0$).}

We also repeat our entire analysis using a few weighting methods, such as directly using the intrinsic weights $\bar{w}_{i}$ and applying the mass-normalized weighting differently, in order to ensure the robustness of our size--mass relation results. In all cases, we find that the qualitative conclusions remain unchanged: SAGA satellites consistently exhibit slightly larger sizes compared to isolated galaxies, and the scatter in the SAGA satellite size--mass relation remains smaller. We also repeat the same analysis using the Tractor catalog sizes for all three samples, and find a qualitatively consistent result. Using Tractor sizes introduces more scatter as a consequence of more outliers in the size--mass relation, and we find an even larger offset in median size between the samples. Additionally, we note that across these tests the only substantial change to our results was that the isolated galaxy sample has an appreciably shallower slope when fit using the weights computed directly from the completeness correction, $\bar{w}_i$, without mass normalization. This can be easily understood by the fact that the isolated sample has the largest completeness correction in terms of galaxy number density, evident in the right panel of Figure~\ref{fig:mass-hists}, so the lower mass end of the size--mass plane is dominating the fit when using $\bar{w}_i$. This is precisely the reason why we compute and use $w_i$: accounting for the increasing stellar mass function toward low-mass enables us to report a size--mass relation robust across a wide mass range. 


\subsection{The Ellipticity and \sersic~Index Distributions Across Environments}
\label{sec:sersic-env}

\begin{figure}[!tbp]
    \centering
    \includegraphics[width=\linewidth,clip]{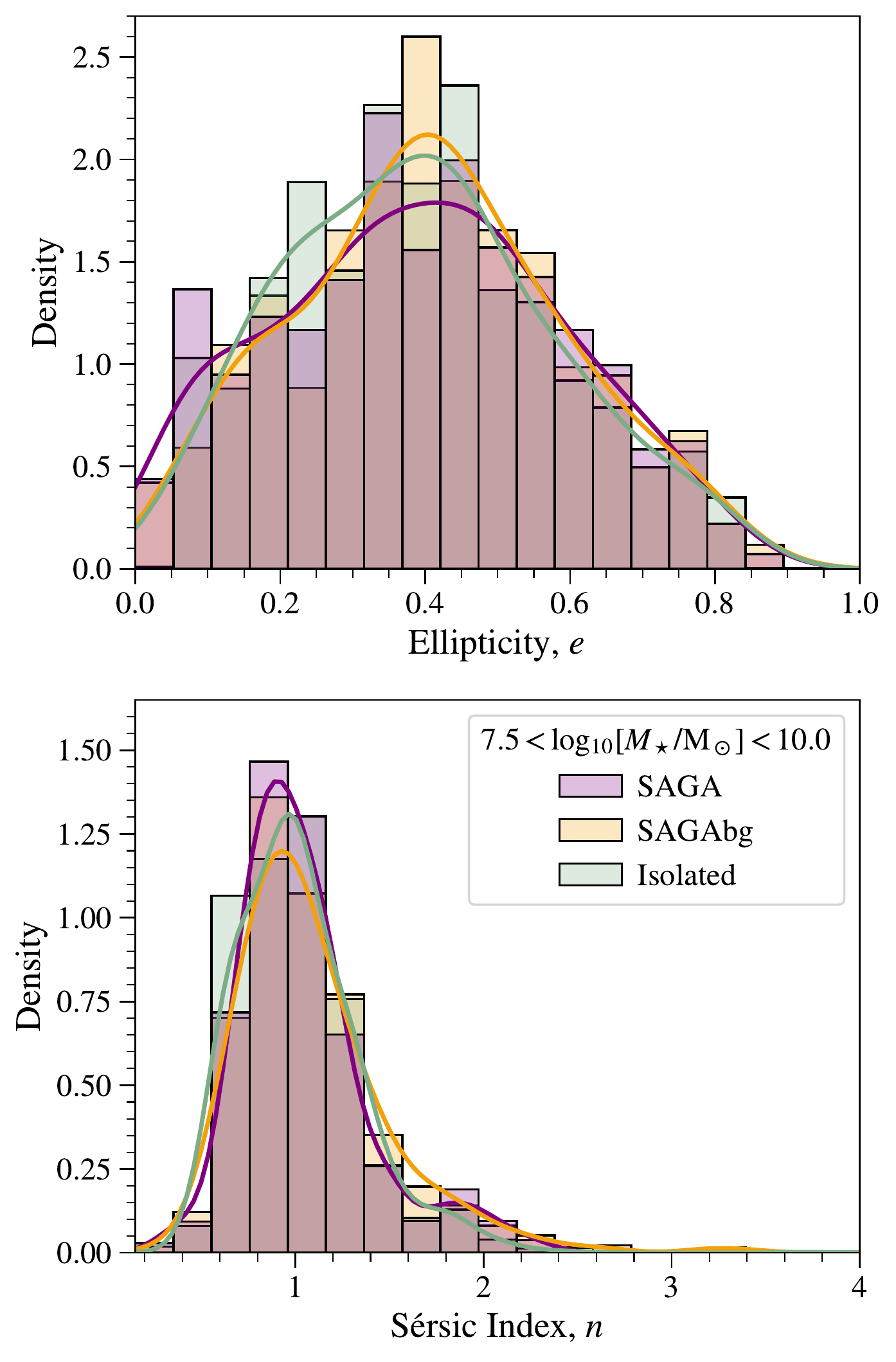}
    \caption{Completeness corrected ellipticity ($e$, \textit{top}) and \sersic~index ($n$, \textit{bottom}) distributions of SAGA satellites (\sagacolor), SAGAbg galaxies (\sbgcolor), and isolated galaxies (\isocolor) in the mass range $7.5 < \logsm < 10.0$. Gaussian KDE density functions are included to illustrate the overall shape of each distribution, and distributions are normalized to density for direct comparison between populations despite sample size difference. There is no evidence for the ellipticity and \sersic~index distributions differing as a function of environment, with the exception of the quenched satellite population.}
    \label{fig:shapes-samples}
\end{figure}

We next investigate \sersic~index and ellipticity across our three samples, and find little difference between the distributions. Figure~\ref{fig:shapes-samples} shows KDE histograms for the ellipticity and \sersic~index distributions for the SAGA satellites, SAGAbg sample, and isolated galaxies. We do not find evidence for significant differences in these structural properties across the samples.   

As discussed in Section~\ref{sec:ellip_ser_saga}, the quenched and star-forming SAGA satellites have distinct ellipticity distributions, with quenched satellites tending toward lower ellipticities. We first examine whether the shapes or light profiles of high-mass quenched galaxies ($\logsm > 9.0$) differ across our samples; we do not find any clear differences, though we note that the number of high-mass quenched galaxies is small in all samples. To directly test if the star-forming satellite population differs from the isolated population, we compare only the star-forming galaxies in each sample. We find no evidence for the star-forming isolated and SAGAbg distributions differing from that of star-forming SAGA satellites. This is illustrated in the bottom panels of Figure~\ref{fig:smr-sf-only}, where the ellipticity and \sersic~index distributions for only the star-forming galaxies in each sample are shown. We again verify that our findings are not driven by differing sample completeness corrections as in Section~\ref{sec:ellip_ser_saga}. 

Both structural properties show consistent distributions across environment, indicating star-forming galaxies have similar stellar light profiles regardless of environment. This suggests that any \sersic~index or ellipticity evolution is largely independent of environment for star-forming galaxies, and these structural properties may be primarily set by internal processes once a galaxy has quenched or by the quenching processes themselves.

\section{Discussion}
\label{sec:discussion}

We have presented our measurements of the sizes and structural properties of SAGA satellites, comparing both quenched and star-forming populations and contrasting them with stellar mass- and redshift-matched environmentally averaged and isolated samples. We now turn to discussing the broader implications of these results in the context of previous work.

\subsection{Comparison with Literature Size--Mass Relations}
\label{sec:struct-lit}

\begin{figure}[!tbp]
    \centering
    \includegraphics[width=\linewidth,clip]{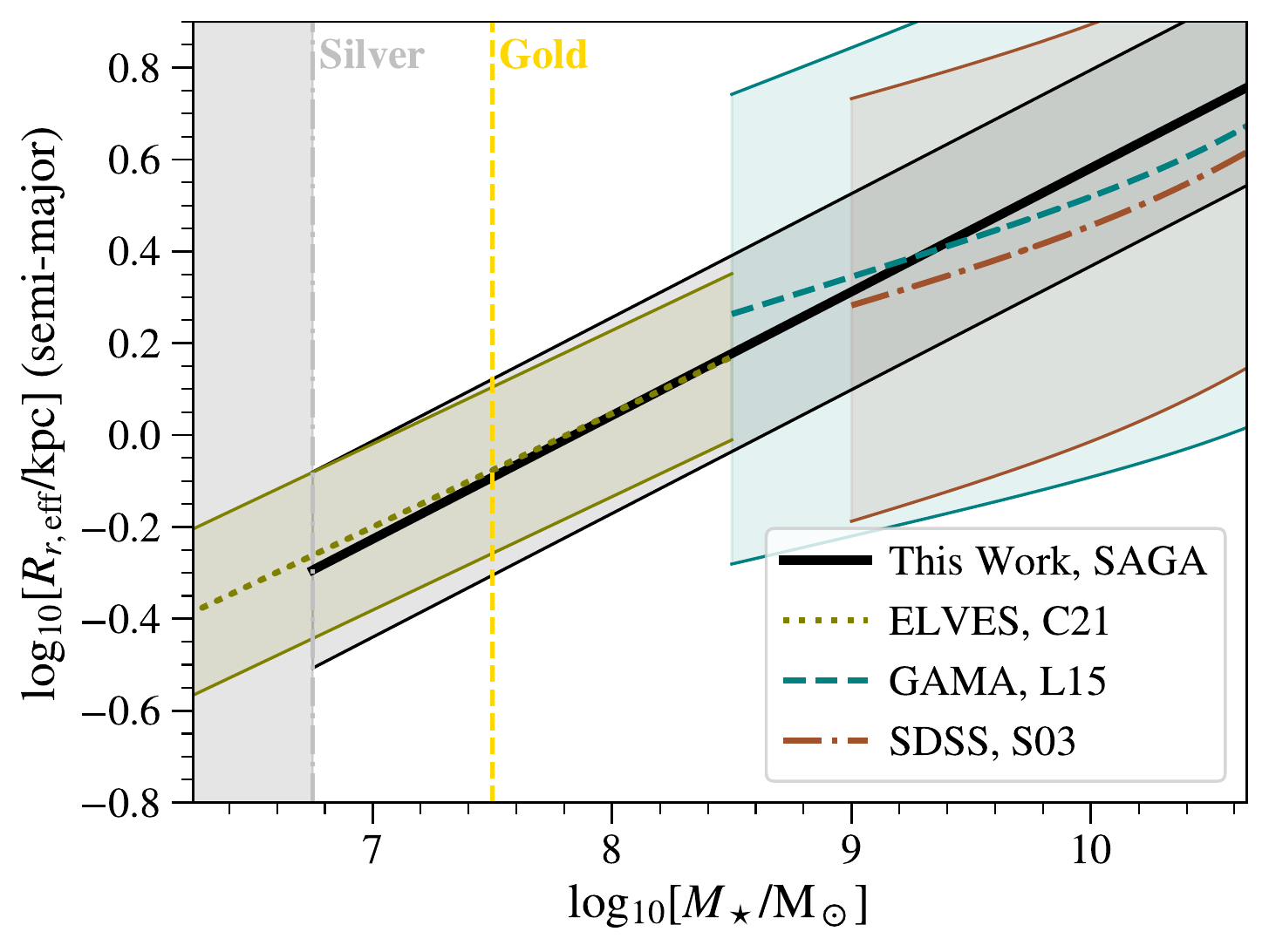}
    \caption{The completeness-corrected size--mass relation for the SAGA \textit{Gold} sample (Equation~\ref{eq:saga-gold-fit}) compared to literature observations. Overall, our relation is largely consistent with the ELVES survey \citep[olive dotted line;][]{carlsten_2021ApJ...922..267C}, GAMA  \citep[teal dashed line;][]{lange_2015MNRAS.447.2603L}, and SDSS \citep[sienna dot dashed line;][]{shen_2003MNRAS.343..978S}, although these relations are determined from different methods and data. ELVES covers a lower mass range, and GAMA and SDSS are both fit to data at higher masses. Shaded regions around the fits represent the reported $1\sigma$ uncertainties.}
    \label{fig:size--mass-lit}
\end{figure}

The size--mass relation we report for SAGA satellites is broadly consistent with results from both observational surveys and simulations. In Figure~\ref{fig:size--mass-lit}, we compare our relation to those of ELVES \citep[olive dotted line;][]{carlsten_2021ApJ...922..267C}, GAMA \citep[teal dashed line;][]{lange_2015MNRAS.447.2603L}, and SDSS \citep[sienna dot dashed line;][]{shen_2003MNRAS.343..978S}. The ELVES survey reported a size--mass relation for satellite galaxies spanning the stellar mass range of $5.5 < \logsm < 8.5$ \citep{carlsten_2021ApJ...922..267C}. While they do not have star formation information, they divide their sample using a color and morphology based criteria into early- and late-type satellites. They find no significant dependence of the size--mass relation on this indicator, although there is tentative evidence late-type satellites have a slightly steeper size--mass relation, consistent with what we find in SAGA for star-forming satellites. We include their full satellite sample fit for comparison in Figure~\ref{fig:size--mass-lit}. 

At higher masses, \citet{lange_2015MNRAS.447.2603L} report multiple size--mass relations from the GAMA survey ($\logsm \gtrsim 8.5$), spanning a larger redshift range ($z<0.1$) than our sample. For comparison, we include their double-power law fit for all galaxies with \sersic~indices $n<2.5$ in Figure~\ref{fig:size--mass-lit}. Despite the differences in mass and redshift range, we find that our results remain broadly consistent with this relation in the regime where our samples overlap. We also compare with their size--mass relations for early- and late-type galaxies, which are reported using several classifications (e.g., morphology, color, \sersic~index). Across these definitions, the qualitative trends they report are consistent with our findings: late-type galaxies tend to be larger in our considered mass range above $\logsm \gtrsim 8.5$, and early-type galaxies exhibit a flatter slope. However, their size-mass relations for both early- and late-type galaxies are shallower than our measured quenched and star-forming satellite relations. Extrapolated to lower mass, the GAMA early- and late-type relations would be expected to intersect around $\logsm \sim 5$, whereas we observe the quenched and star-forming satellite relations intersect around $\logsm \sim 8$. 

Similarly, we compare our results to the size--mass relation from SDSS, as originally reported by \citet{shen_2003MNRAS.343..978S}. This relation was fit for late-type galaxies at stellar masses above $\logsm > 8.8$, using circularized $z$-band sizes. Following \citet{lange_2015MNRAS.447.2603L}, we apply a wavelength-dependent size correction from \citet{kelvin_2012MNRAS.421.1007K}, multiplying the \citet{shen_2003MNRAS.343..978S} relation by an $r$-to-$z$ band size ratio of $1.075$ for late-type galaxies. Despite the differences in methodology, which make a direct comparison challenging, we find that our relation for SAGA satellites remains broadly consistent with the size--mass relation derived from larger surveys at the high-mass end. 

Our SAGA satellite size--mass relation is also consistent with recent results from certain simulations. \citet{klein_2024MNRAS.532..538K} report a size–mass relation for galaxies in the FIRE simulations that closely matches both the slope and normalization of our observed relation, as well as the level of intrinsic scatter. In the TNG simulations, studies have reported a bimodal size distribution among low-mass galaxies ($\logsm > 8.4$), with a dominant population consistent with our observed relation and a secondary population of compact, smaller systems \citep{genel_2018MNRAS.474.3976G, haslbauer_2019A&A...626A..47H}. However, this bimodality is not apparent in observational samples such as GAMA \citep{dealmeida_2024A&A...687A.131D}, nor do we observe it in the SAGA satellite population. 

\subsection{Impact of Environment on Galaxy Sizes}

Numerous studies have explored the effect of environment on the size--mass relation at higher stellar masses ($\logsm \gtrsim 10$), with conflicting results: some report larger galaxy sizes in denser environments  \citep[e.g.,][]{yoon_2017ApJ...834...73Y}, others find smaller sizes \citep[e.g.,][]{cebrian_2014MNRAS.444..682C}, or no significant environmental trend as a function of either local density \citep[e.g.,][]{kauffmann_2004MNRAS.353..713K} or central/satellite distinction \citep[e.g.,][]{spindler_2017MNRAS.468..333S}. 

Even when focusing specifically on low-mass galaxies in the MW-like environment, the picture remains similarly uncertain. \citet{carlsten_2021ApJ...922..267C}, using the ELVES survey, compare MW-like satellites to both cluster and field galaxies.\footnote{ELVES is not perfectly matched to SAGA in terms of local environment: it is a volume-limited sample that spans a wider range in host masses and includes the MW and M31. About half of its hosts (15/31) would not meet SAGA’s MW-analog criteria in luminosity, halo mass, or environment. Additionally, ELVES covers the Local Volume, which represents a different large-scale environment than what is mostly sampled by SAGA. However, for the purposes of discussion we treat it as a comparable dataset.} They observe MW-like satellites tend to have larger sizes than galaxies in the field; however, they note their field sample is incomplete and biased against low-surface-brightness galaxies. \citet{chamba_2024A&A...689A..28C} similarly explored the environment-size correlation using MW-like satellites from both ELVES and SAGA DR2, but find smaller satellite sizes compared to field galaxies. 

Additionally, although the cluster environment differs from what is considered in this work, both \citet{carlsten_2021ApJ...922..267C} and \citet{chamba_2024A&A...689A..28C} examine cluster satellites and reach opposing conclusions: \citet{chamba_2024A&A...689A..28C} find cluster satellites tend to have smaller sizes, while \citet{carlsten_2021ApJ...922..267C} find a small size enhancement ($+0.04$ dex) in cluster satellites relative to MW-like satellites, using half-light sizes.  \citet{chamba_2024A&A...689A..28C} interpret their observed truncated edge sizes in dense environments as a consequence of outside-in quenching driven by ram-pressure stripping and tidal forces, which remove low-density outer HI gas and truncate star formation in the outskirts, preventing extended stellar distributions. By contrast, \citet{carlsten_2021ApJ...922..267C} interpret their observed offset toward larger half-light sizes in dense environments as due to the effects of tidal heating. 

The differing results between these studies may reflect a combination of factors, including variations in size definitions, sample selection criteria, and completeness corrections. Unlike the half-light sizes used in \citet{carlsten_2021ApJ...922..267C}, \citet{chamba_2024A&A...689A..28C} define a size metric that traces the observed galaxy edge (corresponding roughly to a surface brightness of $\sim 26.5$ mag/arcsec$^2$, more comparable but not equivalent to $R_{80}$, the radius containing 80\% of a galaxy's light). Some of the SAGA satellite galaxies included in this work overlap with their sample, but we have independently remeasured structural properties. 

In agreement with \citet{carlsten_2021ApJ...922..267C}, we find that SAGA satellites are systematically larger than our isolated galaxy samples, both in terms of their half-light and 80\%-light sizes, with median offsets between $\sim$10-24\%. This discrepancy is not solely due to the presence of quenched satellites in the SAGA sample, nor is it entirely an observational incompleteness effect. The similarity in \sersic~index distributions across all samples implies that these structural changes do not correspond to drastic differences in morphology, while still resulting in a measurable increase in the size of the stellar distribution. 

These results provide evidence for an environmentally-driven offset in the size--mass relation even in relatively low-density, MW-like halos, but the physical origin of the effect remains unclear. One possibility is a direct impact of environment on stellar structure, through mechanisms such as tidal heating or the proposed ``resonant stripping" mechanism that can inflate low-mass disky galaxies without the need to strip gas \citep{donghia_2009Natur.460..605D}. The observed elevation in the size--mass relation could also possibly be explained by stellar mass loss due to environmental quenching or stripping, which would shift the sample toward lower stellar masses for a given size. 
Alternatively, environment may influence galaxy sizes indirectly by altering dark matter (DM) halo properties. The low-mass end of the stellar-to-halo mass relation itself is expected to have considerable scatter \citep{sawala_2016MNRAS.457.1931S, garrison-kimmel_2017MNRAS.464.3108G, munshi_2021ApJ...923...35M, santos-santos_2022MNRAS.515.3685S, nadler_2023ApJ...945..159N, kim_2024arXiv240815214K} that is not yet entirely constrained observationally \citep{allen_2019MNRAS.488.4916A, nadler_2020ApJ...893...48N, danieli_2023ApJ...956....6D, monzon_2024ApJ...976..197M}. Satellites may also be more DM-depleted than isolated galaxies due to environmental stripping \citep{kaufmann_2007MNRAS.382.1187K, munshi_2017arXiv170506286M}, which could affect their structural scaling relations. Several studies have suggested this indirect scenario to explain possible environment-size correlations. \citet{wang_2020ApJ...889...37W} find that galaxy sizes scale more tightly with halo mass than with stellar mass, and report that size--mass relation differences are driven by variations in halo mass rather than satellite/central classification. 

Recent work by \citet{mercado_2025ApJ...983...93M} has attempted to disentangle these direct and indirect effects of environment on galaxy sizes in FIREbox. Using perturbation index as a proxy for environmental density, they find that galaxies in denser environments are systematically larger at fixed stellar mass, consistent with the results we present here. They suggest environmental factors not only shape the stellar distribution directly, but also indirectly affect the size--mass relation by impacting the inner DM distribution and halo mass. These effects become especially pronounced at lower mass, where galaxies are more easily affected by environment. In this work, we see qualitatively similar trends to what is found in \citet{mercado_2025ApJ...983...93M}, including an increase in the statistical significance of the observed environment-driven size offset toward low stellar masses. 

\subsection{Implications for the Halo Radius Model}

In the halo radius model \citep{kravtsov_2013ApJ...764L..31K}, galaxy size scales linearly with the virial radius of the dark matter halo, implying that halo angular momentum is the dominant factor controlling galaxy sizes. This framework successfully reproduces observed size-dependent clustering in SDSS, where smaller galaxies at fixed stellar mass cluster more strongly than larger ones \citep{hearin_2019MNRAS.489.1805H}. This model implies that satellite galaxies should be smaller than central galaxies of the same stellar mass \citep{hill_2025OJAp....8E..52H}, but our results show that at fixed stellar mass, SAGA satellites are systematically larger than isolated galaxies. This finding at first appears to contradict the halo radius model; however, it is possible environmental effects may alter both galaxy and halo properties. In particular, extensions to the halo radius model that incorporate additional dependencies such as halo concentration can mediate the relationship between size and environment \citep{jiang_2019MNRAS.488.4801J}. Also, we note that our sample probes low-mass galaxies ($7.5 < \logsm < 10.0$), below the regime typically tested in SDSS clustering analyses, but the halo radius model has been shown to predict a half-light radius distribution at low mass that is consistent with MW satellite observations without altering satellite sizes after infall \citep{nadler_2020ApJ...893...48N}. However, our findings suggest that in MW-mass halos, satellite sizes are not simply inherited from their halo properties at infall. Instead, post-infall processes such as stripping or tidal heating may affect satellite sizes. The observed offset we find challenges models in which galaxy size is primarily set by halo properties independent of environment and instead points to additional, environmentally dependent mechanisms that shape either halo properties or satellite structure.

\subsection{Structural Evolution Associated with Satellite Quenching} 

\reply{Environmental processes are known to induce structural evolution in galaxies \citep[e.g.,][]{mayer_2001ApJ...547L.123M, kormendy_2012ApJS..198....2K}. At stellar masses below $\logsm \lesssim 9.0$, environmental processes can also quench star formation via ram-pressure stripping, gravitational interactions (harassment), or starvation \citep{gunn_1972ApJ...176....1G, tonnesen_2007ApJ...671.1434T, tonnesen_2009ApJ...694..789T, boselli_2014A&ARv..22...74B}. For galaxies in MW-like environments, \citet{carlsten_2021ApJ...922..267C} find the structural properties of early- and late-type satellites are remarkably similar, and as a result suggest that quenching does not involve morphological transformation. By contrast, our results suggest that satellites of MW-analogs do undergo structural evolution as they quench.}

We find that quenched and star-forming satellites occupy different distributions in the size--mass plane, with quenched satellites following a shallower size--mass relation. The two relations intersect near $\logsm \sim 8$, such that quenched satellites are smaller than their star-forming counterparts at higher mass and larger at lower mass. At $\logsm \lesssim 8.5$, quenched satellites also show a shift toward lower ellipticity, implying rounder stellar distributions relative to both star-forming satellites and more massive quenched satellites. However, the \sersic~index distributions for star-forming and quenched satellites are nearly indistinguishable, suggesting that their overall light profiles remain similar.

Taken together, these findings point toward a need for mild morphological transformation associated with quenching. \reply{At lower masses ($\logsm \lesssim 8.5$), quenching mechanisms are predicted to inflate the stellar distribution through impulsive heating and suppress rotational support, transforming disky progenitors into more spheroidal systems \citep{mayer_2001ApJ...547L.123M, kazantzidis_2011ApJ...726...98K, watkins_2023MNRAS.521.2012W}. The effectiveness of these transformations depends on the progenitor's internal structure and orbital history, and some simulations show that stellar mass loss becomes significant in the late stages of stripping \citep{penarrubia_2008ApJ...673..226P}. The larger sizes and rounder morphologies of quenched satellites relative to star-forming ones are consistent with these scenarios.}

\reply{At higher stellar masses, satellite trends resemble those of field populations, with star-forming galaxies being larger than quiescent galaxies of the same mass. This may reflect a shift in dominant quenching mechanisms: \citet{fillingham_2015MNRAS.454.2039F} find that satellites above $\logsm \sim 8$ quench primarily through starvation, whereas lower-mass satellites quench more rapidly through stripping.} That being said, at the high-mass end ($\logsm \gtrsim 9$), quenching and thus structural evolution may be driven in part by processes unrelated to the immediate MW-like host. \reply{Quenching may result from group pre-processing prior to infall \citep[e.g.,][]{delucia_2012MNRAS.423.1277D, wetzel_2013MNRAS.432..336W, behroozi_2014ApJ...787..156B} or internal feedback processes \citep[e.g.,][]{elbadry_2016ApJ...820..131E, wheeler_2017MNRAS.465.2420W}. These mechanisms could suppress star formation and alter the structural properties of massive satellites, accounting for the mass-dependent differences in observed quenched satellite properties.} 

Nonetheless, there may still be environmental influence \reply{from the MW-analog host} on the structure of quenched galaxies in this regime, given that we find quenched satellites tend to be slightly larger than quenched isolated galaxies (though the number of isolated quenched galaxies in our sample is small). This is consistent with findings from \citet{yoon_2023ApJ...957...59Y}, who report that low-mass quenched galaxies in high-density environments exhibit larger sizes and shallower mass–size relations than their counterparts in isolated environments. \reply{Additionally, at the high-mass end, the lower ellipticities of star-forming satellites may reflect increasing bulge contributions that reduce the apparent flattening of the light profile, even when disks are still present \citep[e.g.,][]{bluck_2014MNRAS.441..599B}.}

\reply{Many quenching processes are expected to operate more efficiently closer to the host, and \citet{geha_2024ApJ...976..118G} demonstrated that quenched SAGA satellites are more radially concentrated than their star-forming counterparts. If these same processes also drive structural evolution, one might expect normalized size to vary with projected distance from the host. However, we find no clear radial trends in either the star-forming or quenched populations. We cannot further distinguish which physical processes drive the structural differences we observe, but our results nonetheless suggest that some morphological evolution is required to explain the distinct size--mass relations of quenched and star-forming satellites.}

\section{Summary}
\label{sec:conc}

In this work, we leverage the high spectroscopic completeness of the SAGA Survey and homogeneously measure structural properties \reply{using \texttt{pysersic}} to robustly compare galaxy sizes, ellipticities, and brightness profiles across environments. We explore the structural properties of SAGA satellite galaxies, and present new measurements of the size–mass relation for quenched and star-forming satellites. Using mass- and redshift-matched galaxy samples, we constrain the effect of environment on galaxy sizes and structural properties. Our results indicate that, even within the MW-like environment, satellites exhibit systematic structural differences compared to isolated galaxies. Our main results are summarized as follows: 

\begin{enumerate}
    \item We present the size–mass relation for all SAGA satellites above $\logsm > 7.5$ and find a best-fit relation of \reply{$\log_{10}(R_{r, \mathrm{eff}} / \text{kpc}) = 0.27 \logsm - 2.11$}, with a lognormal scatter of \reply{$\sigma = 0.21$} dex (Figures~\ref{fig:size--mass-saga} and~\ref{fig:size--mass-lit}).
    \item We measure the size–mass relations for star-forming and quenched SAGA satellites in the stellar mass range of $6.75 < \logsm < 10$ (Figure~\ref{fig:size--mass-saga}), and find the relation for quenched satellites exhibits a shallower slope than that of star-forming satellites, with the two crossing over around a stellar mass of $\logsm \sim 8$.
    \item As a population, satellite galaxies around MW-like hosts are larger than isolated galaxies by $0.12$ dex or 24\% on average across the stellar mass range $7.5 < \logsm < 10$ (Figure~\ref{fig:size--mass-samples}). This offset is roughly constant across mass, but is most statistically significant at the lowest and highest masses. When restricting to star-forming galaxies only (Figure~\ref{fig:smr-sf-only}), SAGA satellites remain systematically larger than isolated counterparts by 0.12 dex, indicating that the observed offset is not driven by the presence of quenched satellites. 
    \item Quenched SAGA satellites tend toward lower ellipticities than star-forming satellites, with the low-ellipticity population dominated by lower-mass galaxies ($\logsm \lesssim 8.5$, Figure~\ref{fig:shapes-saga}). This suggests that quenched satellites at lower mass are, on average, rounder than their star-forming counterparts. In contrast, the ellipticity distributions of star-forming satellites and isolated star-forming galaxies are consistent with each other. 
    \item The \sersic~index distributions for star-forming and quenched SAGA satellites are statistically similar, and consistent with being drawn from the same distribution (Figure~\ref{fig:shapes-saga}). This result is consistent with previous findings that \sersic~index is not a reliable discriminator between early- and late-type morphologies at low-mass \citep{lange_2015MNRAS.447.2603L}. \reply{Similarly, we find the \sersic~index distributions of satellites and isolated galaxies are statistically indistinguishable.}
    \item We examine alternative measures of galaxy size. We recover the expected wavelength dependence of the size--mass relation and find that $g$- and $r$-band sizes are nearly identical for all samples, while $z$-band sizes are systematically smaller (Figure~\ref{fig:size--mass-r80-bands}), consistent with what is found in other studies \citep[e.g.][]{lange_2015MNRAS.447.2603L}. We also find $R_{80}$ scales uniformly with $R_{\mathrm{eff}}$ across the samples, corresponding to an offset of $\sim$0.25 dex larger (Figure~\ref{fig:size--mass-r80-bands}). The roughly constant offset between $R_{80}$ and $R_{\mathrm{eff}}$ is as expected, given the similarities between the \sersic~index distributions of all samples. 
    \item Our satellite size--mass relation agrees with that of the ELVES Survey \citep{carlsten_2021ApJ...922..267C}, and our observed environmental offset is consistent with predictions from the cosmological volume simulation FIREbox \citep{mercado_2025ApJ...983...93M}. 
\end{enumerate}

Our findings indicate that, despite the many factors affecting galaxy sizes at low mass, environment plays a meaningful structural role, though the precise mechanisms remain unclear. Further modeling work \reply{could help to clarify} which physical processes, such as tidal heating or dark matter stripping, are primarily responsible for producing this effect. \reply{Future observations could also further disentangle these mechanisms; for example, measurements of $V_{\rm rot}/\sigma$ for low-mass galaxies across environments would constrain the extent to which tidal stirring produces dispersion-dominated systems, as opposed to internal feedback or non-environmental processes, and following the predictions of \citet{fillingham_2015MNRAS.454.2039F}.} The magnitude \reply{of the effect of environment on structural properties} may also be better quantified with the onset of large spectroscopic surveys like DESI, which will provide larger low-mass galaxy samples across environments. 


\medskip
\input{acknowledgments}

\software{
    Numpy \citep{numpy, 2020NumPy-Array},
    SciPy \citep{scipy, 2020SciPy-NMeth},
    Matplotlib \citep{matplotlib},
    IPython \citep{ipython},
    Jupyter \citep{jupyter},
    Astropy \citep{astropy}, 
    Photutils \citep{2016ascl.soft09011B},
    SEP \citep{1996A&AS..117..393B, 2016JOSS....1...58B}, 
    pysersic \citep{2023JOSS....8.5703P}, 
    dustmaps \citep{2018JOSS....3..695M}, 
    \reply{pysr \citep{cranmer_2023arXiv230501582C}}
    }

\bibliographystyle{aasjournalv631} 
\bibliography{ref}

\appendix

\section{An accurate approximation for the Sérsic \texorpdfstring{$\lowercase{b}_{\lowercase{n}}$}{b\_n} from \texorpdfstring{$ 0.1 < {\lowercase{n}} < 10$}{0.1 < n < 10}}
\label{sec:app-pysersic}

\reply{ In this work, we present \sersic~fits to galaxy images, which includes sampling over the \sersic~index $n$. This index generally varies between $\sim$1$-$8 in real galaxies; however, low-mass galaxies can exhibit lower \sersic~indices in the range of $n \lesssim 0.6$ \citep{lange_2015MNRAS.447.2603L}. This necessitates sampling below the standard prior lower limit on $n$ of 0.6 imposed in the \texttt{pysersic} code, both because the true value may lie below this threshold, and because truncating the sampler can bias the posterior even when the true \sersic~index is slightly above the cutoff. We thus implemented an update to the \texttt{pysersic} code \citep{2023JOSS....8.5703P} to facilitate sampling below typical default \sersic~cutoffs, as described below. }

\reply{The \sersic~function models the light profile of a galaxy as:}
\begin{equation}
I(R) = I_e \exp\left\{ -b_n \left[ \left( \frac{R}{R_e} \right)^{1/n} - 1 \right] \right\}
\end{equation}
\reply{It contains a term, $b_n(n)$, which satisfies an equation involving the complete and incomplete gamma functions ($\Gamma$,$\gamma$) and depends on $n$. While this equation is trivial to solve for any $n$ on modern computers, it involves finding a root, which makes it intractable for applications where many thousands (or millions) of \sersic~models must be created in minimal time. Several approximating functions have been used in the literature to provide reasonable approximations of $b_n$ \citep{Capaccioli:1989, Ciotti:1999}; however, these functions are relatively simple and diverge from the true values in different $n$ regimes, particularly for $n\lesssim0.36$ and $n>8$.} 

\reply{We use symbolic regression to find a mathematical approximation to $b_n(n)$ that is computationally performant and accurate to within 5 parts in $10^{6}$ for all $n$ between 0.1 and 10, resulting in fully negligible error in effective radius for produced models. Across the galaxy samples used in this work, only 23 galaxies have best fit \sersic~indices below $n<0.36$ with a minimum value of $n=0.27$. Therefore, while classical approximations will work the vast majority of the time, there are rare cases where low-mass galaxies are best fit by low-$n$ models.}

\reply{ We use the package \texttt{pysr} \citep[High-Performance Symbolic Regression in Python and Julia;][]{cranmer_2023arXiv230501582C} to perform the regression. Because the \cite{Ciotti:1999} approximation fits the numerically-derived $b_n$ values over most of the domain to high degrees of accuracy already, we choose to fit instead the residual between this approximation and truth. This makes our final approximation a higher-order correction to the \cite{Ciotti:1999} expression, which attempts to model out inconsistencies between the two, functionally. Practically, the residual being fit is flat and near 0 everywhere except at $n<0.36$, where it rapidly becomes a steep function of $n$. We use most of the defaults defined with the \texttt{pysr} package, allowing binary and unary operators. We also implement some restrictions to avoid nesting functions which trivially reduce (such as a square inside a square root). We use an L2 loss, with a max size parameter of 40 (total complexity) and max depth of 10, limiting the complexity of any particular component. We trigger 64 bit precision for this test and iterated for 1000 steps.}

\begin{figure*}[!tbp]
\centering
  \includegraphics[width=\linewidth,clip]{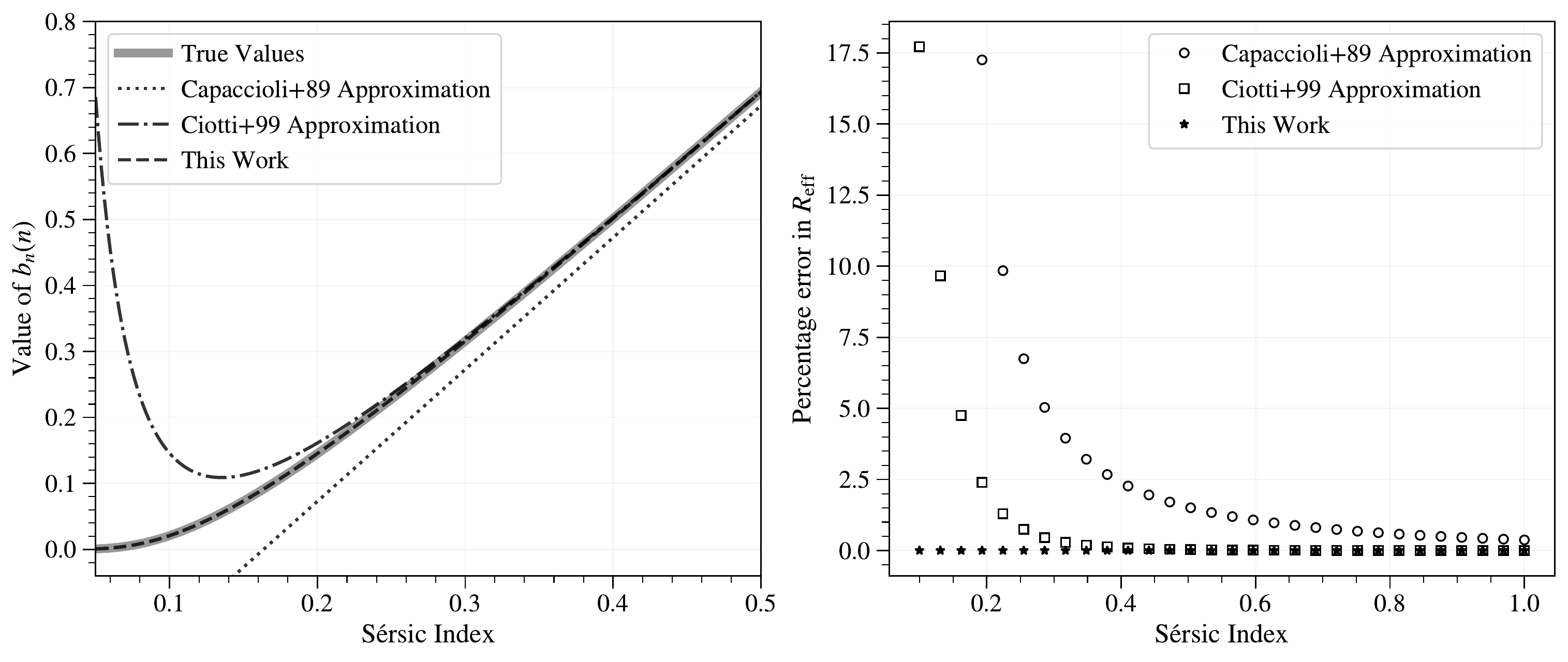}
  \caption{ \reply{ \textit{Left:} Comparison of the \cite{Capaccioli:1989} and \cite{Ciotti:1999} approximations, and the adjusted approximation presented in this work. The new approximation provides a much better fit to the true values for $n<0.36$, while still maintaining high accuracy everywhere else in the domain. \textit{Right:} Percentage error (systematic offset) between input $r_e$ and actual profile $r_e$ for different approximation schemes for $b_n$. These results align with general heuristics known in the field; the \cite{Ciotti:1999} approximation begins producing offset effective radii at around $n\sim0.36$, while the \cite{Capaccioli:1989} profile introduces considerable error at low $n$.} }
\label{fig:sersic_bn_approx}
\end{figure*}

\reply{Our final equation is presented in the form }
\begin{equation}
    b_n(n)\approx A(n) + C(n)
\end{equation}
\reply{where $A(n)$ is the \cite{Ciotti:1999} expansion, and $C(N)$ is the correction term derived from the \texttt{pysr} regression. This equation adds to the complexity of the original (it is not a pure expansion anymore), and significantly improves the accuracy of the $b_n$ prediction. In Equation~\ref{eqn:best_sersic_approx}, we show the full approximation for clarity, including $A(n)$, which is marked with an underbrace. The numerical values of the constants in Equation~\ref{eqn:best_sersic_approx} are presented in Table~\ref{tab:sersic_approx}.}
\vspace{0.5em}
\begin{equation}\label{eqn:best_sersic_approx}
b_n(n) \approx
\underbrace{
    2n - \frac{1}{3} + \frac{4}{405n} + \frac{46}{25515n^2}
}_{A(n)}
+
\left(
\frac{
    \dfrac{a_0}{n + e^n} + a_1
}{
    (a_2 - n)^2
}
\right)
\cdot
\left(
\frac{
    \log n + \dfrac{a_3 n + a_4}{n^{-4} + a_5}
}{
    n + a_6
}
\right)
\end{equation}
\vspace{0.5em}

\begin{table*}[htb]\label{tab:sersic_approx}
\centering
\caption{Fitted Constants for \sersic~\( b_n(n) \) Approximation in Equation~\ref{eqn:best_sersic_approx}}
\begin{tabular}{ll}
\hline
Symbol & Value \\
\hline
 \( a_0 \) & $ 1.8073182821237496\times10^{-4}$ \\ \( a_1 \) & $3.7026973571904255\times10^{-5}$ \\ \( a_2 \) & $-0.09149183119702775$ \\ \( a_3 \) & $2.6248718397705195$\\ \( a_4 \) & $-0.9727511612512357$ \\ \( a_5 \) & $94.78011643586419$ \\ \( a_6 \) & $-0.006044236674273689$\\ 
 \end{tabular}
\end{table*}

\reply{ Figure~\ref{fig:sersic_bn_approx} shows the approximations of \citet{Capaccioli:1989} and \citet{Ciotti:1999} alongside the approximation values from our adjusted function. Despite being broadly applicable over many indices, the equation of \cite{Ciotti:1999} breaks down at $n<0.36$. We find that the new approximation vastly improves the accuracy for $n<0.36$, while not introducing any additional inaccuracy at higher $n$. The maximum absolute deviation from truth in the new approximation is $1.08\times10^{-5}$, occurring at $n=0.055$; for $n>0.1$, the maximum absolute deviation is $5.36\times10^{-6}$ , occurring at $n=0.11$. } \reply{ This new approximation is now the default used in \texttt{pysersic} for deriving $b_n$ for a given model when using the \texttt{PixelRenderer}.} 

\section{Comparison with Legacy Tractor Catalogs}
\label{sec:app-tractor}

\begin{figure}[tbh]
    \centering
    \includegraphics[width=\linewidth,clip]{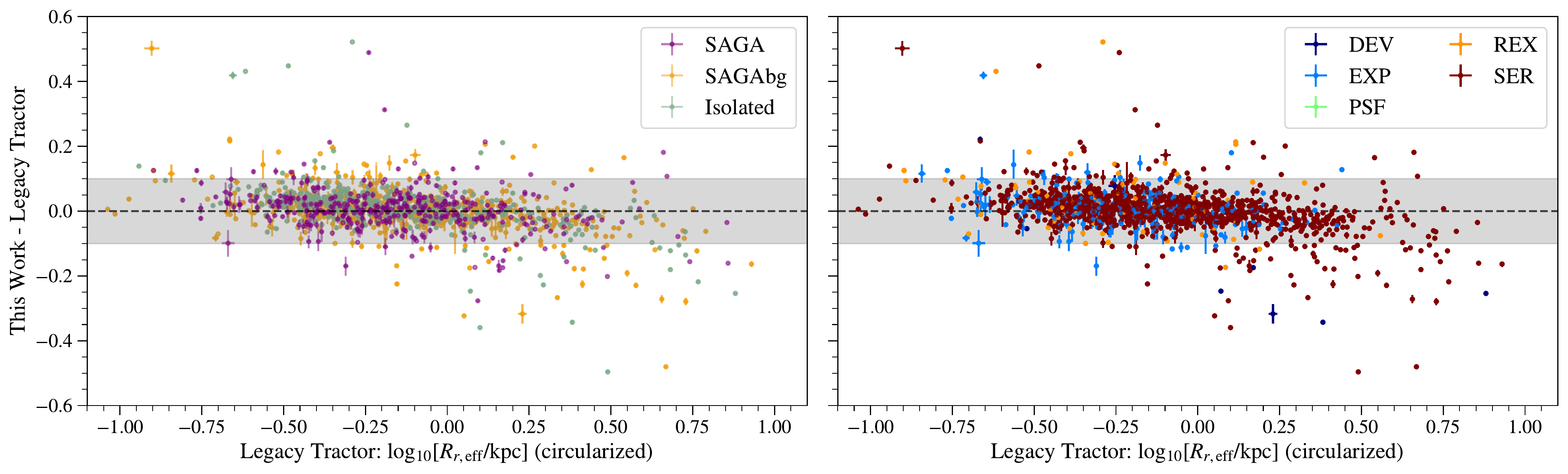}
    \caption{
    Comparison of remeasured sizes with Legacy Tractor catalog sizes. We compare the circularized effective radius ($\log_{10}[R_{\mathrm{eff}}/\mathrm{kpc}]$), plotting the residuals between our measurements and the Tractor catalog values against the Tractor values. 
    The left panel shows residuals colored by sample, while the right panel shows residuals colored by morphological model used in the Tractor fit (\texttt{DEV}, \texttt{EXP}, \texttt{PSF}, \texttt{REX}, and \texttt{SER}). 
    The gray shaded region indicates $\pm 0.1$ dex. The residuals are small for the majority of the data, demonstrating the consistency between our measurements and the Tractor catalog. While there are some systematic differences in size measurements, particularly at the extremes of the size distribution, they do not affect the samples differently. 
    }
    \label{fig:tractor_comparison}
\end{figure}

We compare our remeasured sizes (see Section~\ref{sec:measure-prof}) to the sizes reported in the Legacy DR9 Tractor catalogs \citep{dey_2019AJ....157..168D, lang_2016ascl.soft04008L} to evaluate the consistency of our measurements. Figure~\ref{fig:tractor_comparison} shows the results of this comparison. We compute the circularized effective radius ($R_{\mathrm{eff, circ}}$) measured by our method to compare with the sizes reported in the Legacy Tractor catalogs. The circularized effective radius ($R_{\mathrm{eff, circ}}$) is a measure derived from the semi-major axis ($R_{\mathrm{eff, SMA}}$) and the axial ratio ($b/a$), using the formula $R_{\mathrm{eff, circ}} = R_{\mathrm{eff, SMA}} \cdot \sqrt{b/a}$. The Tractor catalog values are obtained by cross-matching using the X-match service on the NSF's NOIRLab Astro Data Lab platform\footnote{\href{https://datalab.noirlab.edu/index.php}{datalab.noirlab.edu}}, identifying all matches within 5 arcseconds of each galaxy's coordinates. Among these matches, the best match is selected by prioritizing entries with \sersic~model fits. If multiple matches remain, the one with the smallest difference in dereddened $r$-band apparent magnitude is selected. The two panels display the residuals in  $\log_{10}[R_{r, \mathrm{eff}} / \mathrm{kpc}]$ between our measurements and the Tractor values, with points colored by sample on the left and by the morphological type used in the Tractor fit on the right. These models are defined in order of increasing complexity as follows: \texttt{PSF} denotes a point source, \texttt{REX} denotes a round exponential galaxy model, \texttt{DEV} denotes a de Vaucouleurs model, \texttt{EXP} denotes an exponential profile, and \texttt{SER} denotes a \sersic~profile fit. We show residuals as a function of the Tractor catalog sizes.  

The majority of the data are in agreement, with a residual standard deviation of approximately 0.08 dex and a two-sigma range within 0.2 dex. Most discrepancies occur at the extremes of the size distribution. Specifically, our measurements tend to be smaller than the Tractor catalog values for the largest sizes, and larger than the Tractor catalog values for the smallest sizes. At the high-mass end, one common source of disagreement arises from the presence of bars in the galaxies. Bars often lead to systematically larger sizes in the Tractor fits, which are typically associated with a higher axial ratio ($b/a$) value. There are also some outliers ($\sim5$\%) with significant deviations (greater than 0.6 dex), typically resulting from rare catastrophic fits in the Tractor catalog, which are substantially improved through our more careful and detailed fitting procedure outlined in Section~\ref{sec:measure-prof}. 

\section{Galaxy Sizes across the \texorpdfstring{$\lowercase{grz}$}{grz}-bands and Other Size Measures (\texorpdfstring{$R_{80}$}{R80}) }
\label{sec:lambda-r80}

\begin{figure*}[!tbp]
    \centering
    \includegraphics[width=\linewidth,clip]{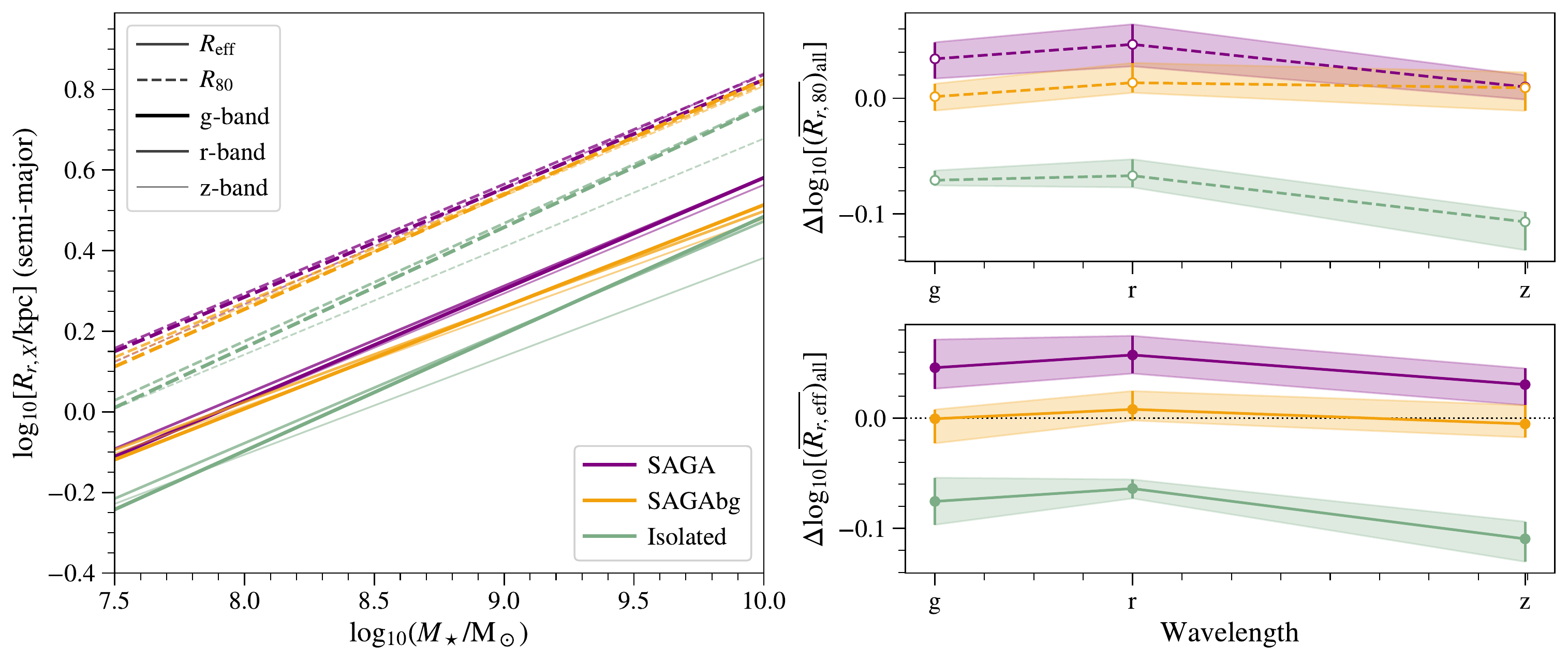}
    \caption{Size–mass relations and wavelength dependence of sizes for all three samples. 
    \textit{Left:} Size–mass relations using both $R_{\mathrm{eff}}$ (solid lines) and $R_{80}$ (dashed lines), measured in the $g$- (thickest), $r$-, and $z$-bands (thinnest). The colors correspond to the three samples: SAGA satellites (\sagacolor), SAGAbg galaxies (\sbgcolor), and isolated galaxies (\isocolor). The offset between $R_{80}$ and $R_{\mathrm{eff}}$ is $\sim$0.25 dex and consistent across samples, reflecting the similarity in \sersic~indices. 
    \textit{Right:} Median galaxy size in each band, normalized to the average size across $g$, $r$, and $z$ for each sample, shown for $R_{80}$ (top) and $R_{\mathrm{eff}}$ (bottom). Sizes in the $g$ and $r$ bands are nearly identical, while $z$-band sizes are systematically smaller. These trends are consistent with previous work showing that galaxy sizes decrease in redder bands \citep[e.g.][]{lange_2015MNRAS.447.2603L}.
    }
    \label{fig:size--mass-r80-bands}
\end{figure*}

In addition to $R_{\mathrm{eff}}$, we also investigate the size–mass relation using $R_{80}$, as this may be a better proxy of total galaxy extent than $R_{\mathrm{eff}}$. The left panel of Figure~\ref{fig:size--mass-r80-bands} illustrates the size--mass relation for all three samples in both size measures: $R_{\mathrm{eff}}$ (solid lines, equivalent to the results presented in Figure~\ref{fig:size--mass-samples}) and $R_{80}$ (dashed lines). We find that the overall normalization of the relation increases by $\sim$0.25 dex from $R_{\mathrm{eff}}$ to $R_{80}$, and this increase in the size--mass relation is equivalent for all three samples. This is expected, as $R_{80}$ differs from $R_{\mathrm{eff}}$ primarily based on \sersic~index, and the majority of galaxies in our samples have $1 < n < 2$ (see Sections~\ref{sec:ellip_ser_saga} and~\ref{sec:sersic-env}). At higher mass, the behavior of the size--mass relation in $R_{80}$ as compared with $R_{\mathrm{eff}}$ is shown to differ most for quiescent galaxies, which tend toward higher \sersic~indices \citep[e.g.][]{miller_2019ApJ...872L..14M}. 

Similarly, we examine the dependence of galaxy sizes on wavelength by comparing $R_{\mathrm{eff}}$ and $R_{80}$ measured in the $g$-, $r$-, and $z$-bands. The left panel of Figure~\ref{fig:size--mass-r80-bands} highlights both the $R_{\mathrm{eff}}$ and $R_{80}$ size--mass relations across all three bands (shown in decreasing linewidth), and the right panel shows the normalized-to-average median size in each of these measures as a function of wavelength. We find that sizes in the $g$ and $r$ bands are nearly identical across all samples, while $z$-band sizes are slightly smaller in most cases, consistent with findings from \citet{lange_2015MNRAS.447.2603L} that smaller galaxy sizes are measured in redder bands. The scale of this wavelength dependence is slightly less pronounced in the $R_{80}$ sizes of the SAGA and SAGAbg samples. This is visible in the top right panel of Figure~\ref{fig:size--mass-r80-bands}, where we can see the median normalized size in the $z$ band is not substantially lower than the $g$ and $r$ bands. However, we only probe a relatively narrow wavelength range across the $g$-, $r$-, and $z$-bands, so we do not constrain the overall wavelength dependence as well as \citet{lange_2015MNRAS.447.2603L}, which measures this relationship between the $g$- and $K_S$-bands, spanning double the wavelength range included here.

Another feature to note is the slightly different slopes of the size--mass relation in these bands. We can see this difference in the left panel of Figure~\ref{fig:size--mass-r80-bands}. In the case of the isolated sample, the $z$-band size--mass relation is shallower, so the wavelength-size dependence is most pronounced at the high-mass end. In contrast, in the case of the SAGA satellite sample, the opposite is true and the $z$-band size--mass relation is steeper, so the wavelength-size dependence is most pronounced at the low-mass end. 

One could imagine this difference reflects how different physical pictures dominate growth for satellites and isolated galaxies. Often, the wavelength-size relation is interpreted to be a consequence of stellar population and metallicity gradients that may be due to inside-out formation, where bluer light is more concentrated toward the center where star-formation is occurring and redder light extends further out due to stellar halos from accreted material \citep[e.g.,][]{lange_2015MNRAS.447.2603L, jia_2024ApJ...977..165J}. This interpretation is complicated by dust attenuation effects which can cause a similar offset in sizes as the level of attenuation varies with wavelength. However, if we assume the wavelength dependence is tracing stellar population differences, the decreasing correlation between size and wavelength towards lower stellar mass for the isolated sample may be indicative of decreasing stellar populations gradients in isolated low-mass systems compared with their higher-mass counterparts. Further, one could imagine interpreting the increased wavelength-size dependence for low-mass satellites to be reflective of some environmental outside-in quenching effect, resulting in a higher concentration of bluer light towards galaxy centers. However, we again caution that this work probes only a narrow range in wavelength, so these trends would need to be confirmed with additional observations spanning a wider range in wavelength. 

\end{document}

%% file: authors.tex
\author[0000-0002-8320-2198]{Yasmeen~Asali}
\affiliation{Department of Astronomy, Yale University, New Haven, CT 06520, USA}
\email[show]{yasmeen.asali@yale.edu}
\correspondingauthor{Yasmeen~Asali}

\author[0000-0002-7007-9725]{Marla~Geha}
\affiliation{Department of Astronomy, Yale University, New Haven, CT 06520, USA}
\email{marla.geha@yale.edu}

\author[0000-0002-0332-177X]{Erin~Kado-Fong}
\affiliation{Department of Physics and Yale Center for Astronomy \& Astrophysics, Yale University, New Haven, CT 06520, USA}
\email{erin.kado-fong@yale.edu}

\author[0000-0002-1200-0820]{Yao-Yuan~Mao}
\affiliation{Department of Physics and Astronomy, University of Utah, Salt Lake City, UT 84112, USA}
\email{yymao@astro.utah.edu}

\author[0000-0003-2229-011X]{Risa~H.~Wechsler}
\affiliation{Kavli Institute for Particle Astrophysics and Cosmology and Department of Physics, Stanford University, Stanford, CA 94305, USA}
\affiliation{SLAC National Accelerator Laboratory, Menlo Park, CA 94025, USA}
\email{rwechsler@stanford.edu}

\author[0000-0002-4739-046X]{Mithi~A.~C.~de~los~Reyes}
\affiliation{Department of Physics and Astronomy, Amherst College, Amherst, MA 01002}
\email{mdelosreyes@amherst.edu}

\author[0000-0002-7075-9931]{Imad~Pasha}
\affiliation{Department of Astronomy, Yale University, New Haven, CT 06520, USA}
\email{imad.pasha@yale.edu}

\author[0000-0002-3204-1742]{Nitya~Kallivayalil}
\affiliation{Department of Astronomy, University of Virginia, Charlottesville, VA 22904, USA}
\email{njk3r@virginia.edu}

\author[0000-0002-1182-3825]{Ethan~O.~Nadler}
\affiliation{Department of Astronomy $\&$ Astrophysics, University of California, San Diego, La Jolla, CA 92093, USA}
\email{enadler@ucsd.edu}

\author[0000-0002-9599-310X]{Erik~J.~Tollerud}
\affiliation{Space Telescope Science Institute, Baltimore, MD 21218, USA}
\email{etollerud@stsci.edu}

\author[0000-0001-8913-626X]{Yunchong~Wang}
\affiliation{Kavli Institute for Particle Astrophysics and Cosmology and Department of Physics, Stanford University, Stanford, CA 94305, USA}
\affiliation{SLAC National Accelerator Laboratory, Menlo Park, CA 94025, USA}
\email{ycwang19@stanford.edu}

\author[0000-0001-6065-7483]{Benjamin~Weiner}
\affiliation{Department of Astronomy and Steward Observatory, University of Arizona, Tucson, AZ 85721, USA}
\email{bjw@as.arizona.edu}

\author[0000-0002-5077-881X]{John~F.~Wu}
\affiliation{Space Telescope Science Institute, Baltimore, MD 21218, USA}
\affiliation{Center for Astrophysical Sciences, Johns Hopkins University, Baltimore, MD 21218, USA}
\affiliation{Department of Computer Science, Johns Hopkins University, Baltimore, MD 21218, USA}
\email{jowu@stsci.edu}

%% file: acknowledgments.tex
The authors would like to thank Kaustav Mitra, Robyn Sanderson, and Francisco Mercado for the many insightful discussions that greatly improved this manuscript. \reply{The authors would also like to thank the anonymous referee for their helpful feedback.} The authors would like to thank Dustin Lang for developing and maintaining the Legacy Surveys Viewer. \reply{YA and IP would like to thank Tim Miller for his input on the updated version of \texttt{pysersic} used in this work.} YA would like to thank Pratik Gandhi, Chris Lindsay, and Tiger Lu for the many discussions and suggestions. YA and MG were supported in part by a grant~from the Howard Hughes Medical Institute (HHMI) through the HHMI Professors Program. 



The Legacy Surveys consist of three individual and complementary projects: the Dark Energy Camera Legacy Survey (DECaLS; Proposal ID \#2014B-0404; PIs: David Schlegel and Arjun Dey), the Beijing-Arizona Sky Survey (BASS; NOAO Prop. ID \#2015A-0801; PIs: Zhou Xu and Xiaohui Fan), and the Mayall z-band Legacy Survey (MzLS; Prop. ID \#2016A-0453; PI: Arjun Dey). DECaLS, BASS and MzLS together include data obtained, respectively, at the Blanco telescope, Cerro Tololo Inter-American Observatory, NSF’s NOIRLab; the Bok telescope, Steward Observatory, University of Arizona; and the Mayall telescope, Kitt Peak National Observatory, NOIRLab. Pipeline processing and analyses of the data were supported by NOIRLab and the Lawrence Berkeley National Laboratory (LBNL). The Legacy Surveys project is honored to be permitted to conduct astronomical research on Iolkam Du’ag (Kitt Peak), a mountain with particular significance to the Tohono O’odham Nation.

NOIRLab is operated by the Association of Universities for Research in Astronomy (AURA) under a cooperative agreement with the National Science Foundation. LBNL is managed by the Regents of the University of California under contract to the U.S. Department of Energy.

This project used data obtained with the Dark Energy Camera (DECam), which was constructed by the Dark Energy Survey (DES) collaboration. Funding for the DES Projects has been provided by the U.S. Department of Energy, the U.S. National Science Foundation, the Ministry of Science and Education of Spain, the Science and Technology Facilities Council of the United Kingdom, the Higher Education Funding Council for England, the National Center for Supercomputing Applications at the University of Illinois at Urbana-Champaign, the Kavli Institute of Cosmological Physics at the University of Chicago, Center for Cosmology and Astro-Particle Physics at the Ohio State University, the Mitchell Institute for Fundamental Physics and Astronomy at Texas A\&M University, Financiadora de Estudos e Projetos, Fundacao Carlos Chagas Filho de Amparo, Financiadora de Estudos e Projetos, Fundacao Carlos Chagas Filho de Amparo a Pesquisa do Estado do Rio de Janeiro, Conselho Nacional de Desenvolvimento Cientifico e Tecnologico and the Ministerio da Ciencia, Tecnologia e Inovacao, the Deutsche Forschungsgemeinschaft and the Collaborating Institutions in the Dark Energy Survey. The Collaborating Institutions are Argonne National Laboratory, the University of California at Santa Cruz, the University of Cambridge, Centro de Investigaciones Energeticas, Medioambientales y Tecnologicas-Madrid, the University of Chicago, University College London, the DES-Brazil Consortium, the University of Edinburgh, the Eidgenossische Technische Hochschule (ETH) Zurich, Fermi National Accelerator Laboratory, the University of Illinois at Urbana-Champaign, the Institut de Ciencies de l’Espai (IEEC/CSIC), the Institut de Fisica d’Altes Energies, Lawrence Berkeley National Laboratory, the Ludwig Maximilians Universitat Munchen and the associated Excellence Cluster Universe, the University of Michigan, NSF’s NOIRLab, the University of Nottingham, the Ohio State University, the University of Pennsylvania, the University of Portsmouth, SLAC National Accelerator Laboratory, Stanford University, the University of Sussex, and Texas A\&M University.

BASS is a key project of the Telescope Access Program (TAP), which has been funded by the National Astronomical Observatories of China, the Chinese Academy of Sciences (the Strategic Priority Research Program “The Emergence of Cosmological Structures” Grant \# XDB09000000), and the Special Fund for Astronomy from the Ministry of Finance. The BASS is also supported by the External Cooperation Program of Chinese Academy of Sciences (Grant \# 114A11KYSB20160057), and Chinese National Natural Science Foundation (Grant \# 12120101003, \# 11433005).

The Legacy Survey team makes use of data products from the Near-Earth Object Wide-field Infrared Survey Explorer (NEOWISE), which is a project of the Jet Propulsion Laboratory/California Institute of Technology. NEOWISE is funded by the National Aeronautics and Space Administration.

The Legacy Surveys imaging of the DESI footprint is supported by the Director, Office of Science, Office of High Energy Physics of the U.S. Department of Energy under Contract No. DE-AC02-05CH1123, by the National Energy Research Scientific Computing Center, a DOE Office of Science User Facility under the same contract; and by the U.S. National Science Foundation, Division of Astronomical Sciences under Contract No. AST-0950945 to NOAO.